%% file: gluino-lsp_munuSSM.tex
\documentclass[12pt]{article}

\usepackage{amsmath}
\usepackage{cancel}
\usepackage{enumitem}
\usepackage{amssymb}
\usepackage{soul}
\usepackage{tabularx}
\usepackage{xcolor}
\usepackage{booktabs}
\usepackage{subcaption} 
\usepackage{colortbl}
\usepackage{authblk}
\usepackage{xspace}
\DeclareUnicodeCharacter{00A0}{ }

\newcolumntype{C}{>{\centering\arraybackslash}X}
\input{defsV2}

\allowdisplaybreaks

\usepackage{feynmp}
\graphicspath{{./Figures/Diagrams/}}
\usepackage{enumitem} 

\usepackage{array, makecell}
\usepackage{boldline}
\usepackage[nosort]{cite}

\title{\bf{Searching for Gluino LSP at the LHC } }

\author[a]{Paulina Knees \thanks{pknees@df.uba.ar}}
\author[b,c]{Essodjolo Kpatcha\thanks{kpatcha.essodjolo@uam.es}}
\author[d]{I\~naki Lara\thanks{inaki.lara@fuw.edu.pl}}
\author[a,e]{Daniel~E.~L\'opez-Fogliani\thanks{daniel.lopez@df.uba.ar}}
\author[b,c]{Carlos~Mu\~noz\thanks{c.munoz@uam.es}} 
\author[f]{Natsumi Nagata\thanks{natsumi@hep-th.phys.s.u-tokyo.ac.jp}}
\author[g]{Hidetoshi Otono\thanks{otono@phys.kyushu-u.ac.jp}}

\affil[a]{Instituto de F\'isica de Buenos Aires UBA \& CONICET, Departamento de F\'isica, Facultad de Ciencia Exactas y Naturales, Universidad de Buenos Aires, 1428 Buenos Aires, Argentina}
\affil[b]{Departamento de F\'{\i}sica Te\'{o}rica, Universidad Aut\'{o}noma de Madrid (UAM), Campus de Cantoblanco, 28049 Madrid, Spain}
\affil[c]{Instituto de F\'{\i}sica Te\'{o}rica (IFT) UAM-CSIC, Campus de Cantoblanco, 28049 Madrid, Spain}
\affil[d]{Faculty of Physics, University of Warsaw, Pasteura 5, 02-093 Warsaw, Poland}
\affil[e]{{Pontificia Universidad Católica Argentina, Av. Alicia Moreau de Justo 1500, 1107~Buenos~Aires, Argentina}}
\affil[f] {Department of Physics, University of Tokyo, Tokyo 113-0033, Japan}
\affil[g] {Department of Physics, Kyushu University, Fukuoka 819-0395, Japan}

\date{}

\begin{document}


\maketitle

\begin{abstract}
We analyse relevant signals expected at the LHC, assuming that the gluino is the lightest supersymmetric particle (LSP) in the framework of the $\mu\nu$SSM.
In this $R$-parity violating model, the presence of couplings involving right-handed neutrinos solves simultaneously the $\mu$ problem and the accommodations of neutrino masses and mixing angles. We study
gluino pair production in quark-antiquark and gluon-gluon collisions. The main decay channels for the gluino LSP are the three-body decays to two quarks and a lepton or a neutrino. In both cases, the leading channels occur for the third family of quarks. We compare the predictions of this scenario with LHC searches for prompt and long-lived particles.To analyse the parameter space we sample the $\mu\nu$SSM for a gluino LSP, paying special attention to reproduce the current experimental data on neutrino and Higgs physics, as well as flavour observables. Our results imply a lower limit on the mass of the gluino LSP of about 2600 GeV, and an upper limit for the decay length of about 6 cm.

\end{abstract}



\clearpage 

\tableofcontents 

\section{Introduction}
\label{sec:intro}
The `$\mu$~from~$\nu$' Supersymmetric Standard Model
($\mn$)~\cite{LopezFogliani:2005yw,Escudero:2008jg} (for a review, see Ref.~\cite{Lopez-Fogliani:2020gzo}) is a predictive model alternative to the 
Minimal Supersymmetric Standard Model (MSSM)~\cite{Nilles:1983ge,Barbieri:1987xf,Haber:1984rc,Gunion:1984yn, Martin:1997ns} 
and the Next-to-MSSM (NMSSM)~\cite{Maniatis:2009re,Ellwanger:2009dp}.
In the $\mn$, the presence of couplings involving right-handed (RH) neutrinos solves simultaneously the $\mu$ problem~\cite{Kim:1983dt,Bae:2019dgg} and the $\nu$-problem (the generation of neutrino masses), without
the need to introduce additional energy scales beyond the supersymmetry (SUSY)-breaking scale. In contrast to the MSSM, and the NMSSM, $R$-parity
and lepton number are not conserved,
leading to a completely different
phenomenology characterized by distinct prompt or displaced
decays of the lightest supersymmetric particle (LSP),
producing multi-leptons/jets/photons with small/moderate missing transverse energy (MET) from 
neutrinos~\cite{Ghosh:2017yeh,Lara:2018rwv,Lara:2018zvf,Kpatcha:2019gmq,Kpatcha:2019pve,Heinemeyer:2021opc,Kpatcha:2021nap,Knees:2023fel}.
The smallness of neutrino masses is directly related with the low decay width of the LSP. Actually, it is also related to the existence of possible candidates for decaying dark matter in the model.
This is the case of 
the gravitino~\cite{Choi:2009ng,GomezVargas:2011ph,Albert:2014hwa,GomezVargas:2017,Gomez-Vargas:2019mqk}, or the axino~\cite{Gomez-Vargas:2019vci}, with lifetimes greater than the age of the Universe.
Recently, the role of sterile neutrinos as decaying dark matter candidates in the framework of the $\mn$ has also being studied~\cite{Knees:2022wbt}. 
It is also worth mentioning, concerning cosmology, that baryon asymmetry might be realized in the
$\mn$ through electroweak (EW) baryogenesis~\cite{Chung:2010cd}. 
The EW sector of the $\mn$ can also explain~\cite{Kpatcha:2019pve,Heinemeyer:2021opc}
the longstanding
discrepancy between the experimental result for the anomalous magnetic
moment of the muon~\cite{Abi:2021gix,Albahri:2021ixb} and its SM prediction~\cite{Aoyama:2020ynm}.\footnote{
In this work we will not try to explain it,
since we are interested in the analysis of a gluino LSP through the decoupling of the rest of the SUSY spectrum.}

Because of $R$-parity violation (RPV) in the $\mn$, basically all SUSY particles are candidates for the LSP, and therefore analyses of the LHC phenomenology associated to each candidate are necessary to test them.
This crucial task, given the current experimental results on SUSY searches, has been mainly concentrated on the EW sector of the $\mn$, analysing left sneutrinos, the right smuon and the bino as candidates for the LSP~\cite{Ghosh:2017yeh,Lara:2018rwv,Lara:2018zvf,Kpatcha:2019gmq,Kpatcha:2019pve,Heinemeyer:2021opc}.
More recently, the colour sector of the $\mn$ has been analysed.
In particular, in Ref.~\cite{Kpatcha:2021nap}
the SUSY partners of the top quark as LSP candidates, i.e. the left and right stops, were considered, whereas in  Ref.~\cite{Knees:2023fel} the right sbottom was also considered. 
The aim of this work is to continue with the systematic analysis of the possible LSP's of the $\mn$, focusing now on the gluino as the LSP.

Thus, we will study the constraints on the parameter space of the model by sampling it to get the gluino as the LSP in a wide range of masses. We will pay special attention
to reproduce 
{neutrino masses and mixing angles~\cite{Capozzi:2017ipn,deSalas:2017kay,deSalas:2018bym,Esteban:2018azc,deSalas:2020pgw,Esteban:2020cvm}.} 
In addition, we will impose on the resulting parameters agreement with Higgs data as well as with flavour observables.

The paper is organized as follows. In Sec.~\ref{sec:model}, we will review the $\mn$ and its relevant parameters for our analysis of neutrino, neutral Higgs and gluino sectors.
In Sec.~\ref{sec:gluinopheno},
we will introduce the phenomenology of the gluino LSP, studying especially its signals at the LHC. They consist mainly of displaced vertices with two quarks and a lepton or a neutrino.
In Sec.~\ref{strategy}, we will discuss the strategy that we will employ to
perform scans searching for points of the parameter space of our scenario compatible with current experimental data on neutrino and Higgs physics, as well as flavour observables such as 
$B$ and $\mu$ decays.
The results of these scans will be presented 
in Sec.~\ref{sec:results}, and applied to show the 
current reach of the LHC search on the parameter space of the gluino LSP based on ATLAS and CMS results~\cite{ATLAS:2022pib, ATLAS:2023tbg, CMS:2019qjk, ATLAS:2017tny, CMS:2020iwv,  ATLAS:2020xyo, ATLAS:2020wjh, ATLAS:2021fbt, ATLAS:2023afl, CMS:2022wjc, CMS:2024trg}.
Finally, 
our conclusions are left for Sec.~\ref{sec:conclusion}.


\section{The $\mu\nu$SSM }
\label{sec:model}
In the $\mn$~\cite{LopezFogliani:2005yw,Escudero:2008jg,Lopez-Fogliani:2020gzo}, the particle content of the MSSM
is extended by RH neutrino superfields $\hat \nu^c_i$. 
{The simplest superpotential of the model 
is the following~\cite{LopezFogliani:2005yw,Escudero:2008jg,Ghosh:2017yeh}: 
\bea
W &=&
\epsilon_{ab} \left(
Y_{e_{ij}}
\, \hat H_d^a\, \hat L^b_i \, \hat e_j^c +
Y_{d_{ij}} 
\, 
\hat H_d^a\, \hat Q^{b}_{i} \, \hat d_{j}^{c} 
+
Y_{u_{ij}} 
\, 
\hat H_u^b\, \hat Q^{a}
\, \hat u_{j}^{c}
\right)
\nonumber\\
&+& 
\epsilon_{ab} \left(
Y_{{\nu}_{ij}} 
\, \hat H_u^b\, \hat L^a_i \, \hat \nu^c_j
-
\lambda_i \, \hat \nu^c_i\, \hat H_u^b \hat H_d^a
\right)
+
\frac{1}{3}
\kappa_{ijk}
\hat \nu^c_i\hat \nu^c_j\hat \nu^c_k,
\label{superpotential}
\eea
where the summation convention is implied on repeated indices, with $i,j,k=1,2,3$ the usual family indices of the SM 
and $a,b=1,2$ $SU(2)_L$ indices with $\epsilon_{ab}$ the totally antisymmetric tensor, $\epsilon_{12}= 1$. 
}

{Working in the framework of a typical low-energy SUSY, the Lagrangian  containing the soft SUSY-breaking terms related to $W$ 
is given by:
\bea
-\mathcal{L}_{\text{soft}}  =&&
\epsilon_{ab} \left(
T_{e_{ij}} \, H_d^a  \, \widetilde L^b_{iL}  \, \widetilde e_{jR}^* +
T_{d_{ij}} \, H_d^a\,   \widetilde Q^b_{iL} \, \widetilde d_{jR}^{*} 
+
T_{u_{ij}} \,  H_u^b \widetilde Q^a_{iL} \widetilde u_{jR}^*
+ \text{h.c.}
\right)
\nonumber \\
&+&
\epsilon_{ab} \left(
T_{{\nu}_{ij}} \, H_u^b \, \widetilde L^a_{iL} \widetilde \nu_{jR}^*
- 
T_{{\lambda}_{i}} \, \widetilde \nu_{iR}^*
\, H_d^a  H_u^b
+ \frac{1}{3} T_{{\kappa}_{ijk}} \, \widetilde \nu_{iR}^*
\widetilde \nu_{jR}^*
\widetilde \nu_{kR}^*
\
+ \text{h.c.}\right)
\nonumber\\
&+&   
m_{\widetilde{Q}_{ijL}}^2
\widetilde{Q}_{iL}^{a*}
\widetilde{Q}^a_{jL}
{+}
m_{\widetilde{u}_{ijR}}^{2}
\widetilde{u}_{iR}^*
\widetilde u_{jR}
+ 
m_{\widetilde{d}_{ijR}}^2
\widetilde{d}_{iR}^*
\widetilde d_{jR}
+
m_{\widetilde{L}_{ijL}}^2
\widetilde{L}_{iL}^{a*}  
\widetilde{L}^a_{jL}
\nonumber\\
&+&
m_{\widetilde{\nu}_{ijR}}^2
\widetilde{\nu}_{iR}^*
\widetilde\nu_{jR} 
+
m_{\widetilde{e}_{ijR}}^2
\widetilde{e}_{iR}^*
\widetilde e_{jR}
+ 
m_{H_d}^2 {H^a_d}^*
H^a_d + m_{H_u}^2 {H^a_u}^*
H^a_u
\nonumber \\
&+&  \frac{1}{2}\, \left(M_3\, {\widetilde g}\, {\widetilde g}
+
M_2\, {\widetilde{W}}\, {\widetilde{W}}
+M_1\, {\widetilde B}^0 \, {\widetilde B}^0 + \text{h.c.} \right).
\label{2:Vsoft}
\eea

In the early universe, not only the EW symmetry is broken, but in addition to the neutral components of the Higgs doublet fields $H_d$ and $H_u$ also the left and right sneutrinos $\widetilde\nu_{iL}$ and $\widetilde\nu_{iR}$
acquire a vacuum expectation value (VEV). {With the choice of CP conservation, they develop real VEVs denoted by:  
\begin{eqnarray}
\langle H_{d}^0\rangle = \frac{v_{d}}{\sqrt 2},\quad 
\langle H_{u}^0\rangle = \frac{v_{u}}{\sqrt 2},\quad 
\langle \widetilde \nu_{iR}\rangle = \frac{v_{iR}}{\sqrt 2},\quad 
\langle \widetilde \nu_{iL}\rangle = \frac{v_{iL}}{\sqrt 2}.
\end{eqnarray}
The EW symmetry breaking is induced by the soft SUSY-breaking terms
producing
$v_{iR}\sim {\order{1 \tev}}$ as a consequence of the right sneutrino minimization equations in the scalar potential~\cite{LopezFogliani:2005yw,Escudero:2008jg,Ghosh:2017yeh}.
Since $\widetilde\nu_{iR}$ are gauge-singlet fields,
the $\mu$-problem can be solved in total analogy to the
NMSSM
through the presence in the superpotential (\ref{superpotential}) of the trilinear 
terms $\lambda_{i} \, \hat \nu^c_i\,\hat H_u \hat H_d$.
Then, the value of the effective $\mu$-parameter is given by 
$\mu=
\la_i v_{iR}/\sqrt 2$.
These trilinear terms also relate the origin of the $\mu$-term to the origin of neutrino masses and mixing angles, since neutrino Yukawa couplings 
$Y_{{\nu}_{ij}} \hat H_u\, \hat L_i \, \hat \nu^c_j$
 are present in the superpotential {generating Dirac masses for neutrinos, 
$m_{{\mathcal{D}_{ij}}}\equiv Y_{{\nu}_{ij}} {v_u}/{\sqrt 2}$.}
Remarkably, in the $\mu\nu$SSM it is possible to accommodate neutrino masses
and mixing in agreement with experiments~\cite{Capozzi:2017ipn,deSalas:2017kay,deSalas:2018bym,Esteban:2018azc,deSalas:2020pgw,Esteban:2020cvm} via an EW seesaw
mechanism dynamically generated during the EW symmetry breaking~\cite{LopezFogliani:2005yw,Escudero:2008jg,Ghosh:2008yh,Bartl:2009an,Fidalgo:2009dm,Ghosh:2010zi,Liebler:2011tp}. The latter takes place 
through the couplings
$\kappa{_{ijk}} \hat \nu^c_i\hat \nu^c_j\hat \nu^c_k$,
giving rise to effective Majorana masses for RH neutrinos
${\mathcal M}_{ij}
= {2}\kappa_{ijk} {v_{kR}}/{\sqrt 2}$.
Actually, this is possible at tree level even with diagonal Yukawa couplings~\cite{Ghosh:2008yh,Fidalgo:2009dm}.
It is worth noticing here that the neutrino Yukawas discussed above also generate the effective bilinear terms $\mu_i \hat H_u\, \hat L_i $
with $\mu_i=Y_{{\nu}_{ij}} {v_{jR}}/{\sqrt 2}$,
used in the bilinear RPV model (BRPV)~\cite{Barbier:2004ez}.

We conclude, therefore, that the $\mn$ solves not only the
$\mu$-problem, but also the $\nu$-problem, without
the need to introduce energy scales beyond the SUSY-breaking one.

The parameter space of the $\mn$, and in particular the
neutrino, neutral Higgs and gluino sectors are 
relevant for our analysis in order to reproduce neutrino and Higgs data, and to obtain in the spectrum a gluino as the LSP.
In particular, neutrino and Higgs sectors were discussed in Refs.~\cite{Kpatcha:2019gmq,Kpatcha:2019qsz,Kpatcha:2019pve,Heinemeyer:2021opc}, and we refer the reader to those works for details, although we will summarize the results below. First, we discuss here several simplifications that are convenient to take into account given the large number of parameters of the model.
{Using diagonal mass matrices for the scalar fermions, in order to avoid the
strong upper bounds upon the intergenerational scalar mixing (see e.g. Ref.~\cite{Gabbiani:1996hi}), from the eight minimization conditions with respect to $v_d$, $v_u$,
$v_{iR}$ and $v_{i}$ to facilitate the computation we prefer to eliminate
the
soft masses $m_{H_{d}}^{2}$, $m_{H_{u}}^{2}$,  
$m_{\widetilde{\nu}_{iR}}^2$ and
$m_{\widetilde{L}_{iL}}^2$
in favour
of the VEVs.
Also, we assume} for simplicity in what follows the flavour-independent couplings and VEVs $\lambda_i = \lambda$,
$\kappa_{ijk}=\kappa \delta_{ij}\delta_{jk}$, and $v_{iR}= v_{R}$. Then, the Higgsino mass parameter $\mu$, bilinear couplings $\mu_i$ and Dirac and Majorana masses discussed above are given by:
\bea
\mu=3\la \frac{v_{R}}{\sqrt 2}, \;\;\;\;
\mu_i=Y_{{\nu}_{i}}  \frac{v_{R}}{\sqrt 2}, \;\;\;\;
m_{{\mathcal{D}_i}}= Y_{{\nu}_{i}} 
\frac{v_u}{\sqrt 2}, \;\;\;\;
{\mathcal M}
={2}\kappa \frac{v_{R}}{\sqrt 2},
\label{mu2}    
\eea
where 
we have already used the possibility of having diagonal neutrino Yukawa couplings $Y_{{\nu}_{ij}}=Y_{{\nu}_{i}}\delta_{ij}$ in the $\mn$ in order to reproduce neutrino physics.

\subsection{The Neutrino Sector}
\label{neutrino}

For light neutrinos, under the above assumptions, one can obtain
the following simplified formula for the effective mass matrix~\cite{Fidalgo:2009dm}:
\begin{eqnarray}
\label{Limit no mixing Higgsinos gauginos}
(m_{\nu})_{ij} 
\approx
\frac{m_{{\mathcal{D}_i}} m_{{\mathcal{D}_j}} }
{3{\mathcal{M}}}
                   \left(1-3 \delta_{ij}\right)
                   -\frac{v_{iL}v_{jL}}
                   {4M}, \;\;\;\;\;\;\;\;
        \frac{1}{M} \equiv \frac{g'^2}{M_1} + \frac{g^2}{M_2},         
\label{neutrinoph2}
  \end{eqnarray}     
where $g'$, $g$ are the EW gauge couplings, and $M_1$, $M_2$ the bino and wino soft {SUSY-breaking masses}, respectively.
This expression arises from the generalized EW seesaw of the $\mn$, where due to RPV the neutral fermions have the flavour composition
$(\nu_{iL},\widetilde B^0,\widetilde W^0,\widetilde H_{d}^0,\widetilde H_{u}^0,\nu_{iR})$.
The first two terms in Eq.~(\ref{neutrinoph2})
are generated through the mixing 
of $\nu_{iL}$ with 
$\nu_{iR}$-Higgsinos, and the third one 
also include the mixing with the gauginos.
These are the so-called $\nu_{R}$-Higgsino seesaw and gaugino seesaw, respectively~\cite{Fidalgo:2009dm}.
One can see from this equation that {once ${\mathcal M}$ is fixed, as will be done in the parameter analysis of Sec.~\ref{sec:parameter},
the most crucial independent parameters determining {neutrino physics} are}:
\bea
Y_{\nu_i}, \, v_{iL}, \, M_1, \, M_2.
\label{freeparameters}
\eea
Note that this EW scale seesaw implies $Y_{\nu_i}\lsim 10^{-6}$
driving $v_{iL}$ to small values because of the proportional contributions to
$Y_{\nu_i}$ appearing in their minimization equations. {A rough} estimation gives
$v_{iL}\lsim m_{{\mathcal{D}_i}}\lsim 10^{-4}$.

{Considering the normal ordering for the neutrino mass spectrum,
and taking advantage of the 
dominance of the gaugino seesaw for some of the three neutrino families, three
representative type of solutions for neutrino physics using diagonal neutrino Yukawas were obtained in 
Ref.~\cite{Kpatcha:2019gmq}.
In our analysis we will use the so-called type 2 solutions, which have the structure
\bea
M>0, \, \text{with}\,  Y_{\nu_3} < Y_{\nu_1} < Y_{\nu_2}, \, \text{and} \, v_{1L}<v_{2L}\sim v_{3L},
\label{neutrinomassess}
\eea
In this case of type 2, it is easy to find solutions with the gaugino seesaw as the dominant one for the third family. Then, $v_{3L}$ determines the corresponding neutrino mass and $Y_{\nu_3}$ can be small.
On the other hand, the normal ordering for neutrinos determines that the first family dominates the lightest mass eigenstate implying that $Y_{\nu_{1}}< Y_{\nu_{2}}$ and $v_{1L} < v_{2L},v_{3L}$, {with both $\nu_{R}$-Higgsino and gaugino seesaws contributing significantly to the masses of the first and second family}. Taking also into account that the composition of the second and third families in the second mass eigenstate is similar, we expect $v_{2L} \sim v_{3L}$. 
In Ref.~\cite{Kpatcha:2019gmq}, a quantitative analysis of the neutrino sector was carried out, with the result that the hierarchy qualitatively discussed above for Yukawas and VEVs works properly. 
See in particular Fig.~4 of Ref.~\cite{Kpatcha:2019gmq}, where
$\delta m^2=m^2_2-m^2_1$ 
versus $Y_{\nu_{i}}$ and $v_{iL}$ is shown for the scans carried out in that work, using
the results for normal ordering from Ref.~\cite{Esteban:2020cvm}. 

We will argue in Sec.~\ref{sec:results} that the 
other two type of solutions of normal ordering for neutrino physics are not going to modify our results.
The same conclusion is obtained in the case of working with the inverted 
ordering for the neutrino mass spectrum. The structure of the solutions is more involved for this case, because the two heaviest eigenstates are close in mass and the lightest of them has a dominant contribution from the first family. Thus, to choose $Y_{\nu_1}$ as the largest of the neutrino Yukawas helps to satisfy these relations. For the second and third family, a delicate balance between the contributions of $\nu_{R}$-Higgsino and gaugino seesaws is needed in order to obtain the correct mixing angles. In particular, a representative type of solutions for the case of inverted ordering has the structure $M>0$, with
$Y_{\nu_3} \sim Y_{\nu_2} < Y_{\nu_1}$, and $v_{1L}<v_{2L}\sim v_{3L}$.

\subsection{The Higgs Sector}
\label{sec:higgs}

The neutral Higgses are mixed with right and left sneutrinos, since
the neutral scalars and pseudoscalars in the $\mn$ have the flavour composition
$(H_{d}^0, H_{u}^0, \widetilde\nu_{iR}, \widetilde\nu_{iL}) $.
Nevertheless, the left sneutrinos are basically decoupled from the other states, since
the off-diagonal terms of the mass matrix are suppressed by the small $Y_{\nu_{ij}}$ and $v_{iL}$.
Unlike the latter states, the other neutral scalars can be substantially mixed.
Neglecting this mixing between 
the doublet-like Higgses and the three right sneutrinos, the expression of the tree-level mass of the { SM-like Higgs} is~\cite{Escudero:2008jg}:
\begin{eqnarray}
m_h^2 \approx 
m^2_Z \left(\cos^2 2\beta + 10.9\
{\lambda}^2 \sin^2 2\beta\right),
\end{eqnarray}
where $\tan\beta= v_u/v_d$, and $m_Z$ denotes the mass of the $Z$~boson.
Effects lowering (raising) this mass appear when the SM-like Higgs mixes with heavier (lighter) right sneutrinos. The one-loop corrections are basically determined by 
the third-generation soft {SUSY-breaking} parameters $m_{\widetilde u_{3R}}$, $m_{\widetilde Q_{3L}}$ and $T_{u_3}$
(where we have assumed for simplicity that for all soft trilinear parameters
$T_{ij}=T_{i}\delta_{ij}$).
These three parameters, together with the coupling $\lambda$ and $\tan\beta$, are the crucial ones for Higgs physics. 
Their values can ensure that the model contains a scalar boson with a mass around $\sim 125 \gev$ and properties similar to the ones of the SM Higgs boson~\cite{Biekotter:2017xmf,Biekotter:2019gtq,Kpatcha:2019qsz,Biekotter:2020ehh}.

In addition, $\ka$, $v_R$ and the trilinear parameter $T_{\kappa}$ in the 
soft Lagrangian~(\ref{2:Vsoft}),
are the key ingredients to determine
the mass scale of the {right sneutrinos}~\cite{Escudero:2008jg,Ghosh:2008yh}.
For example, for $\lambda\lsim 0.01$ they are basically free from any doublet admixture, and using their minimization equations in the scalar potential
the scalar and pseudoscalar masses can be approximated respectively by~\cite{Ghosh:2014ida,Ghosh:2017yeh}:
\bea
m^2_{\widetilde{\nu}^{\mathcal{R}}_{iR}} \approx   \frac{v_R}{\sqrt 2}
\left(T_{\kappa} + \frac{v_R}{\sqrt 2}\ 4\kappa^2 \right), \quad
m^2_{\widetilde{\nu}^{\mathcal{I}}_{iR}}\approx  - \frac{v_R}{\sqrt 2}\ 3T_{\kappa}.
\label{sps-approx2}
\eea

Finally, $\lambda$ and the trilinear parameter $T_{\lambda}$ 
not only contribute to these masses
for larger values of $\lambda$, but
also control the mixing between the singlet and the doublet states and hence, they contribute in determining their mass scales as discussed in detail in Ref.~\cite{Kpatcha:2019qsz}.
We conclude that the relevant parameters
in the {Higgs (-right sneutrino) sector} are:
\bea
\lambda, \, \kappa, \, \tan\beta, \, v_R, \, T_\kappa, \, T_\lambda, \, T_{u_3}, \,  m_{\widetilde u_{3R}},
\, m_{\widetilde Q_{3L}}.
\label{freeparameterss}
\eea
Note
that the most crucial parameters for the neutrino sector~(\ref{freeparameters}) are basically decoupled from these parameters controlling Higgs physics.
This simplifies the analysis of the parameter space of the model,
as will be discussed in 
Sec.~\ref{sec:parameter}.

\subsection{The Gluino Sector}

The gluino sector of the $\mn$ coincide with that of the MSSM.
Gluinos are the fermionic superpartners of the gluons and, therefore, carry colour charge, thus participating in strong interactions. 
The gluino mass $m_{\tilde{g}}$ arises from the soft SUSY-breaking terms, being essentially determined by the soft mass parameter $M_3$,
and further influenced by loop corrections.
In this work, we will take $M_3$ to be a low-energy free parameter, compute $m_{\tilde{g}}$ taking into account the one loop corrections, and discuss the limits on it from available experimental data.

\bigskip

\noindent
In our analysis of 
Sec.~\ref{sec:results}., 
we will sample the relevant parameter space of the $\mn$, which contains the independent parameters determining neutrino and Higgs physics in Eqs.~(\ref{freeparameters}) and~(\ref{freeparameterss}). Nevertheless, the parameters for neutrino physics $Y_{\nu_i}$, $v_{iL}$, $M_1$ and $M_2$ are essentially decoupled from the parameters 
controlling Higgs physics, as already mentioned.
Thus, for a suitable choice of the former parameters
reproducing neutrino physics, there is still enough freedom to reproduce in addition Higgs data by playing with $\lambda$, $\kappa$, $v_R$, $\tan\beta$, $T_{u_3}$, etc., 
as shown in Refs.~\cite{Kpatcha:2019gmq,Kpatcha:2019pve,Heinemeyer:2021opc}. 
As a consequence, we will not need to scan over most of the latter parameters, relaxing 
our computing task. For this task
we have employed the 
{\tt Multinest}~\cite{Feroz:2008xx} algorithm as optimizer. To compute
the spectrum and the observables we have used {\tt SARAH}~\cite{Staub:2013tta} to generate a 
{\tt SPheno}~\cite{Porod:2003um,Porod:2011nf} version for the model.

\section{Gluino LSP Phenomenology}
\label{sec:gluinopheno}

Gluinos can be abundantly produced at the LHC (see Refs.~\cite{PhysRevD.31.1581,Beenakker:1996ch}). Since they interact only via strong interactions, they can be generated in pairs through quark-antiquark and gluon-gluon collisions, or singly in association with squarks. Gluino-squark associated production can be important for constraining scenarios of nearly degenerate gluinos and squarks, specially for full degeneracy in the squark sector. However, the points studied in the present work have mass differences between the gluino and the lightest squark in the range of hundreds of GeV, therefore the gluino-squark production is subleading with respect to gluino pair production.
Then, we considered pair production as our case of interest, and the cross-sections used for this analysis in Sec.~\ref{sec:lhc} were obtained at NNLO+NNLL 
using NNLL-fast\cite{Beenakker:2016lwe}.\footnote{
The updated version published recently~\cite{Beenakker:2024jwh}, does not modify the results presented in this work.}
In particular, for our range of interest of gluino masses between 500 GeV and 3000 GeV,
the production cross-section is in the range
between $\sim 33.5$ pb and $6\times 10^{-6}$ pb.

\subsection{Decay Modes}
\label{subsec:Decaymodes}
Gluino decays to Standard Model particles have been discussed in the framework of the MSSM (see, e.g., Refs.~\cite{Barbieri:1987ed,Chigusa:2021glp}). In the $\mn$,
there are three relevant decay modes of the gluino LSP.
On the one hand, the gluino has two three-body decays to two quarks and a lepton or a neutrino, mediated by virtual squarks. These are shown in Fig~\ref{fig:decay-chan} left and centre. 
On the other hand, the gluino can decay to gluon and neutrino through a virtual quark-squark loop. This channel is shown in Fig~\ref{fig:decay-chan} right.
However, given the large values of the gluino masses analysed, this channel is subleading compared to the three-body decays.

 \begin{figure}[t!]
     \centering
        \includegraphics[width=0.3\linewidth]{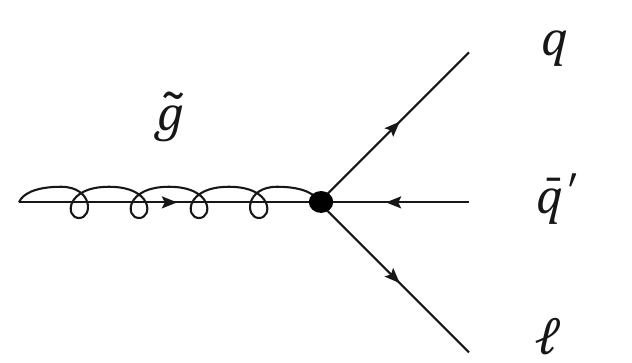}
     \includegraphics[width=0.3\linewidth]{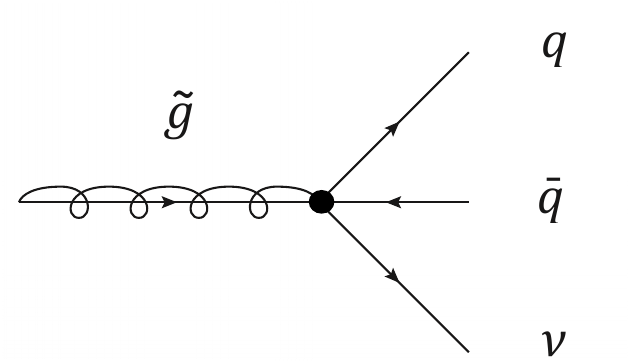}
       \includegraphics[width=0.3\linewidth]{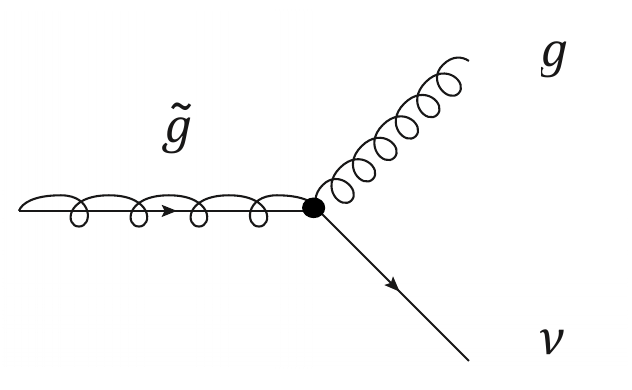}
      \caption{Relevant decay channels in the $\mn$ for a gluino LSP. (left) Decay to quark-antiquark and lepton; (centre) Decay to quark-antiquark and neutrino; (right) Decay to gluon and neutrino. 
      }
     \label{fig:decay-chan}
 \end{figure}

The relevant interactions for the two dominant channels are then given in Appendix~\ref{appendixb}. 
The gluino decays to a quark and a virtual squark, which subsequently decays to another quark and a lepton or a neutrino. Thus, gluino-quark-squark interactions are shown in (\ref{gluinodown}) and
(\ref{gluinoup}), whereas squark-quark-lepton/neutrino interactions are shown 
in (\ref{leptons})-(\ref{neutrinos down}). 

In the latter interactions, one can identify the most important contributions for the squark decays. 
There,
$U^V$ is the matrix which diagonalizes the mass matrix for the  neutral fermions. The entries 
$U^V_{i4}$, $U^V_{i5}$ and {$U^V_{i6}$}, correspond to the mixing between neutrinos and bino, winos and neutral Higgsino $\widetilde H^0_d$, respectively. They can be approximated as  
\begin{eqnarray}
U^V_{i4}\approx\frac{{-}g'}{M_1}\sum_{l}{\frac{v_{lL}}{\sqrt{2}}U^{\text{\tiny{PMNS}}}_{il}},\quad \quad
U^V_{i5}\approx\frac{g}{M_2}\sum_{l}{\frac{v_{lL}}{\sqrt 2}U^{\text{\tiny{PMNS}}}_{il}}, \quad \quad {U^V_{i6}\approx\ \frac{1}{\mu}\sum_l \frac{\mu_l}{\sqrt{2}} U^{\text{\tiny{PMNS}}}_{il}},
\label{--sneutrino-decay-width-2nus2}
\end{eqnarray}
where $U^{\text{\tiny{PMNS}}}_{il}$ are the entries of the PMNS matrix, with
$i$ and $l$ neutrino physical and flavour indices, respectively. 
 We also approximate other entries of the matrices involved in the computation, as follows:  $U_{L, i 4}^{e} \approx 0$ (vanishing charged wino-lepton mixing), $U_{R, i 4}^{e}\approx g v_{iL} /M_2 \sqrt 2$ , $U_{i 7}^{V}\approx 0$ (vanishing 
 neutrino-Higgsino $\widetilde H^0_u$ mixing),
 $U_{L, j 5}^{e}\approx {Y_{e_{i}}v_{iL} }/{ \sqrt 2 \mu}$.

Then,
on the one hand, there are contributions of a squark decaying to quark and lepton 
corresponding to the term multiplying the projector $P_{L}$ and the
second term multiplying the projector $P_{R}$ in both (\ref{leptons}) and
(\ref{eq:appen-ql}).
Thus, they occur mainly through the Yukawa couplings $Y_{t,b}$ of $\widetilde t$ ($\widetilde b$)
with $b$ ($t$) and charged Higgsinos, via the mixing between the latter and $\ell$.
{In addition, there is the contribution of the first term multiplying the projector $P_{R}$ in (\ref{eq:appen-ql}), which occurs through the gauge coupling $g$ of
$\widetilde b$ with $t$ and charged winos, via the mixing between the latter and $\ell$.}
On the other hand, there are contributions of a squark decaying to quark and neutrino corresponding to the first term multiplying the projector $P_{L}$ and the
second and third terms multiplying the projector $P_{R}$ in both (\ref{neutrinos up}) and
(\ref{neutrinos down}).
They occur through 
the gauge couplings $g$ and $g'$ of $\widetilde t$ ($\widetilde b$)
with $t$ ($b$) and neutral winos and binos, via the mixing between the latter and $\nu$.
In addition, there are the contributions of the 
first term multiplying the projector $P_{R}$ and the second term multiplying the projector $P_{L}$ in (\ref{neutrinos down}), which
occur through 
the Yukawa coupling $Y_{b}$ 
of $\widetilde b$
with $b$ and neutral Higgsinos, via the mixing between the latter and $\nu$.
The contributions occurring through the Yukawa coupling $Y_t$ of $\widetilde{t}$ with t and neutral Higgsinos $\widetilde H^0_u$ 
in (\ref{neutrinos up}), via the mixing between the latter and $\nu$,
are neglected below in Eq.~(\ref{gamma-ttnu}) because such a mixing is very small, $U_{i 7}^{V}\approx 0$, as discussed above.

Assuming pure squark states with a common mass fixed at $m_{\widetilde{q}}$, and taking into account the above approximations, the values of the partial decay widths of the gluino LSP can be written as:

\begin{eqnarray}
\sum_{i}\Gamma(\widetilde{g} \to t \bar{t} \nu_i) &\sim& \frac{g^2_s}{3072\pi^3}\frac{m^5_{\widetilde{g}}}{m^4_{\widetilde{q}}}\sum_{i} \left(\frac{1}{2} g^2 {U_{i5}^V}^2+\frac{17}{18}g'^2{U_{i4}^V}^2+\frac{gg'}{3}{U_{i5}^V}{U_{i4}^V} 
\right),
   \label{gamma-ttnu}
   \\
\sum_{i,q\neq t}\Gamma(\widetilde{g} \to q \bar{q} \nu_i) &\sim& \frac{g^2_s}{3072\pi^3}\frac{m^5_{\widetilde{g}}}{m^4_{\widetilde{q}}}\sum_{i} \left(\frac{5}{2} g^2 {U_{i5}^V}^2+\frac{49}{18}g'^2{U_{i4}^V}^2-\frac{gg'}{3}{U_{i5}^V}{U_{i4}^V}+Y^2_b {U_{i6}^V}^2 \right)
   \label{gamma-qqnu}
\\
\Gamma(\widetilde{g} \to t \bar{b} \ell_i^- )&=&\Gamma(\widetilde{g} \to \bar{t} b \bar{\ell}_i^+ ) \sim \frac{g^2_s }{3072\pi^3}\frac{m^5_{\widetilde{g}}}{m^4_{\widetilde{q}}} \left(\frac{g^4}{M^2_2}{\frac{v^2_{iL}}{2}}+\frac{Y^2_tY^2_{e_{i}}v^2_{iL} }{ \mu^2}+\frac{ 2Y^2_b \mu^2_i }{ \mu^2}\right),
   \label{gamma-tbl}
\\
\sum_{q\neq t}\Gamma(\widetilde{g} \to q {\bar q'} \ell_i^- )&=&\sum_{q\neq t}\Gamma(\widetilde{g} \to q' {\bar q} \bar{\ell}_i^+ ) \sim \frac{g^2_s g^2}{3072\pi^3}\frac{m^5_{\widetilde{g}}}{m^4_{\widetilde{q}}} \left(\frac{2 g^4}{M^2_2}{\frac{v^2_{iL}}{2}}\right ),
\label{gamma-qql}
\\
\sum_{i}\Gamma(\widetilde{g} \to g \nu_i) &\sim& \frac{g^4_s Y_t^2\sum_{i} U_{i4}^V }{2048\pi^5}\frac{ m^2_{t} m^3_{\widetilde{g}}}{m^4_{\widetilde{q}}} \left(1+\log \frac{m^2_t}{m^2_{\widetilde{q}}}\right)^2,
\label{gamma-gnu}
\end{eqnarray}
{where in Eq.~(\ref{gamma-tbl}) 
the summation convention on repeated indices does not apply in the second term.} 
{Let us remark nevertheless that the results of Sec.~\ref{sec:results}
have been obtained using the full tree-level numerical computation of decay widths implemented in {\tt SPheno}.}

As can be deduced from these equations, 
when the quarks in the final state belong to the third family, the decay width becomes more relevant
because of the contributions of bottom and top Yukawa couplings in Eqs.~(\ref{gamma-qqnu}) and~(\ref{gamma-tbl}). The exception is the $\widetilde{g} \to t \bar{t} \nu_i$ channel, since the contribution of the top Yukawa coupling is negligible, as discussed above.

\subsection{LHC Searches}
\label{sec:lhc}

The gluino LSP would produce a diverse phenomenology at the LHC, as shown as in Fig.~\ref{fig:decay-chan}. The possible decays include: the production of a lepton ($e$, $\mu$ or $\tau$)/neutrino and 2 quarks (including bottom and top quarks) or the production of a neutrino and a gluino. Consequently, the production of a pair of gluinos will lead to events that can include missing energy, leptons, several jets and $b$-jets. In addition, the decay length of the gluino LSP ranges from mm scale up to $\sim 100$ m. Therefore, different LHC searches will have the highest sensitivity for each case. We will classify the signals according to the lifetime scale and the LHC detection strategy. All constraints will be finally applied to each point analysed.

\pagebreak
\bigskip

\noindent
{\bf Case i) {Large ionization energy loss}}

\vspace{0.2cm}

\noindent 

Whenever the gluino LSP is long-lived enough to measure ionization energy loss and/or time of flight, the searches for heavy charged long-lived particles are able to put strong constraints which are independent of the decay of the gluino LSP. In particular, for each point of the parameter space with a gluino LSP with proper lifetime greater than 3 m, the ATLAS search for heavy charged long-lived particles excludes gluino masses below 2200 GeV~\cite{ATLAS:2022pib,ATLAS:2023tbg}. 
For each point tested, we exclude it if the mass of the gluino LSP is below the exclusion limit obtained by the ATLAS search for the specific value of $c\tau$.

\bigskip

\noindent
{\bf Case ii) {Non-prompt jets}}
\vspace{0.2cm}

\noindent 
The timing capabilities of the CMS electromagnetic calorimeter 
 allow discriminating jets arriving at times significantly larger than the travelling times expected for light hadrons, which are moving at velocities close to the speed of light. This time delay can be associated with two effects: First, the larger indirect path formed by the initial trajectory of a long-lived particle plus the subsequent trajectories of the child particles. Secondly, the slower velocity of the long-lived particle due to the high mass compared to light hadrons. Such analysis is performed by the CMS collaboration in the work~\cite{CMS:2019qjk} in the context of long-lived gluinos decaying to gluons and stable gravitinos, excluding gluinos with masses of $\sim2500$ GeV for lifetimes of $\sim1$ m.

The case where the gluino LSP decays producing a neutrino and a gluon with proper decay lengths above $\sim30$ cm will produce the same signal as the one analysed in~\cite{CMS:2019qjk}. For each point tested in this search, we compare the 95\% observed upper limit on cross-section 
corresponding to events containing a delayed jet for a given parent particle mass and $c\tau$, with the prediction of the signal cross-section calculated as
$\sigma(pp\to\tilde{g}\tilde{g}) \times [BR(\tilde{g}\to g \nu)^2 + 2 BR(\tilde{g}\to g \nu) (1-BR(\tilde{g}\to g \nu))]$.

\bigskip

\noindent 
{\bf Case iii) {Displaced vertices}}
\vspace{0.2cm}

\noindent 
For shorter lifetimes, one can confront the points of the model with the limits from events
with displaced vertices, including jets.
The ATLAS search~\cite{ATLAS:2017tny} targets final states with at least one displaced vertex (DV) with a high reconstructed mass
and a large track multiplicity in events with large missing transverse momentum. It searches in particular for the signal of long-lived gluinos decaying as $\tilde{g}\to \bar{q} q \widetilde{\chi}^0$.
The CMS search~\cite{CMS:2020iwv} also searches for long-lived gluinos decaying into various final-state topologies containing displaced jets. In particular, it searches for the characteristic signals of long-lived gluino pair production decaying as $\tilde{g}\to g\widetilde{\chi}^0$ or  $\tilde{g}\to    \bar{q} q \widetilde{\chi}^0$.

These searches originally target long-lived massive particles with lifetimes in the range $1-10000$ mm. Thus, this search can be sensitive to the gluino LSP when $c\tau$ is in this range. Some of the signal topologies analysed in the search match exactly the ones originated from the decays shown in Fig.~\ref{fig:decay-chan}. The case where the gluino LSP decays producing a neutrino and a gluon, or a pair of quarks and a neutrino,  with proper decay lengths above $\sim1$ mm is contrasted with the model-dependent limits obtained in the ATLAS and CMS searches for displaced vertices. For each point, we compare the 95\% observed upper limit on cross-section
with the prediction of the signal cross-section calculated 
in our model, similarly to Case (ii).

The CMS search~\cite{CMS:2024trg} targets similar topologies and proper decay lengths as does~\cite{CMS:2020iwv}, but using a machine learning algorithm to improve the background rejection power. However, the limits obtained for the topologies studied here are equivalent in both searches, thus no improved limit can be obtained for the gluino LSP from this analysis.

There are other decay channels of the gluino LSP that can produce displaced vertices but do not match exactly the topologies explored in the searches~\cite{ATLAS:2017tny,CMS:2020iwv}. To have a reasonable estimate of the exclusion power of those searches over the additional signals, we use a recast version of the ATLAS analysis within 
CheckMATE-LLP~\cite{Desai:2021jsa}. CheckMATE~\cite{Dercks:2016npn,Kim:2015wza} is a universal tool for the recasting of LHC searches in the context of arbitrary new
physics models. It uses the fast detector simulation framework Delphes~\cite{deFavereau:2013fsa} with customized ATLAS detector card and additional built-in tuning for a more accurate reproduction of experimental efficiencies. The validation of the recast search is discussed in~\cite{Desai:2021jsa}. 
We generate signal Monte Carlo (MC) samples of gluino pair production with MadGraph5\_aMC@NLO-v3.4.2~\cite{Alwall:2014hca,Alwall:2007fs,Alwall:2008qv} at leading order (LO). The hard event corresponds to tree-level production of gluino pairs and includes the emission
of up to two additional partons, 
the NNPDF23LO ~\cite{Ball:2012cx,Buckley:2014ana} PDF set is used. Simulated signal events were passed to 
Pythia-8.306~\cite{Sjostrand:2014zea} for parton showering (PS) and hadronization. Jet matching and merging to parton-shower calculations is accomplished by the MLM algorithm~\cite{Mangano:2006rw}.
Gluino pair-production nominal cross-sections are derived at NNLO+NNLL using  NNLL-fast-3.0~\cite{Beenakker:2016lwe,Beenakker:1997ut,Beenakker:2010nq,Beenakker:2016gmf}.
Finally, we process the events generated through CheckMATE. The results are used to calculate the efficiency ($\epsilon$) of the search, defined as the number of events predicted in the signal region divided by the total number of generated events.

We generate samples for values of the mass equal to $[2000,2250,2500,2750]$ GeV,  $c\tau$ equal to $[1,2,5,10,20,50,100,200,500]$ mm and for all the different combinations of decays shown in Fig.~\ref{fig:decay-chan}, and calculate $\epsilon$ for each case. 
For each point tested in this work, we calculate the $\epsilon$ interpolating from each channel and value of mass and $c\tau$, within the set of $\epsilon$ obtained as described above.

Finally, the point is considered excluded if the total number of events predicted in the signal region of the search\cite{ATLAS:2017tny}, calculated as the sum 
 of $\mathcal{L}\times\sigma(pp\to\tilde{g}\tilde{g})\times BR_{channel}\times \epsilon_{channel}$ over all channels, is greater than the 95\% upper limit on signal events, which correspond to approximately 3 events.

\bigskip

\noindent 
{\bf Case iv) {Displaced jets+muon}}
\vspace{0.2cm}

\noindent 
The ATLAS search for long-lived massive particles in events with a DV and a muon with large impact parameter~\cite{ATLAS:2020xyo}, is designed to look for long-lived stops decaying to a quark and a muon. A similar event topology is possible when the gluino LSP decays producing two quarks and a muon, but in this case producing a vertex with a muon and two jets instead of one. Since there is no explicit restriction in the upper number of tracks matched to the secondary vertex, there is no reason to consider that the additional jet will spoil the efficiency of the search. On the contrary, the requirements on the \emph{full DV selection} of $n^{DV}_{tracks}>3$ and $m_{DV}>20$ GeV will be easier to satisfy with increased hadronic activity associated to the DV.

For each point, we compare the 95\% observed upper limit on cross-section, corresponding to the signal of one DV and one displaced muon originated from the decay of two long-lived stops for a given mass and $c\tau$, with the prediction of the signal cross-section calculated as $\sigma(pp\to\tilde{g}\tilde{g})\times BR(\tilde{g}\to \bar{q} q' \mu)^2$. 

\bigskip

\noindent 
{\bf Case v) {Displaced leptons}}
\vspace{0.2cm}

\noindent 
The gluino LSP can decay producing leptons directly when it decays as $\widetilde{g}\to \bar{q}q' l$ as shown in Fig.~\ref{fig:decay-chan}, or in the decay of a top quark when any of the quarks produced is a top quark. The ATLAS search for displaced leptons~\cite{ATLAS:2020wjh} looks for events with at least two leptonic tracks with an impact parameter greater than 3 mm and, as no requirements were made on missing energy, displaced vertices, or jets, the limits are applicable in general to any BSM model producing high-$p_T$ displaced leptons.

Following a similar procedure to the case of DVs, we recast the analysis within \\ CheckMATE-LLP. The validation of the recast search is discussed in Appendix~\ref{appendix}.  We calculate the $\epsilon$ of the search, following the procedure described for the case of DVs, for the same set of masses and proper decay lengths. 
For each point tested, we calculate the $\epsilon$ interpolating from each channel and value of mass and $c\tau$, within the set of $\epsilon$ obtained as described above.

Finally, the point is considered excluded if the total number of events predicted in the signal region of the search\cite{ATLAS:2020wjh}, calculated as the sum 
 of $\mathcal{L}\times\sigma(pp\to\tilde{g}\tilde{g})\times BR_{channel}\times \epsilon_{channel}$ over all channels, is greater than the 95\% upper limit on signal events, which correspond to 3 events.

\bigskip

\noindent 
{\bf Case vi) {Prompt decays}}
\vspace{0.2cm}

\noindent 
Whenever the proper decay length of the gluino LSP is sufficiently small, $c\tau \lesssim0.5$ mm, the limits calculated for the gluino decaying immediately after production will be sensitive to the gluino LSP. In particular, the ATLAS search~\cite{ATLAS:2021fbt} looks for gluinos decaying as $\tilde{g}\to \bar{q}q\bar{q}q \nu/l$ through the $\lambda'\mathrm{\bar{L}Qd}$ $R$-parity violating coupling. A similar signal is obtained for the gluino LSP from the diagrams in Fig.~\ref{fig:decay-chan} (left and centre).
We constrain the points analysed in this search with $c\tau \lesssim0.5$ mm the 95\% observed upper limit on cross-section, corresponding to the signal of two gluinos decaying as $\tilde{g}\to \bar{q}q\bar{q}q \nu/l$, for massless neutralinos, with the prediction of the signal cross-section calculated as $\sigma(pp\to\tilde{g}\tilde{g})\times BR(\tilde{g}\to \bar{q} q \nu/l)^2$. 

{The ATLAS search~\cite{ATLAS:2023afl} target some of the signals constrained in~\cite{ATLAS:2021fbt} but the former is optimized for signals including at least tree leptons or two same-sign leptons while the latter is optimized for signals with a single isolated lepton. The search~\cite{ATLAS:2023afl} impose constraints over the signal topology described in this section, however the limits obtained are equivalent to  the ones obtained in~\cite{ATLAS:2021fbt}, therefore is not possible to use this search to strengthen the limits over the gluino LSP.}

Finally, The CMS search~\cite{CMS:2022wjc} uses a Deep-Neural-Network based tagger to discriminate decays of long-lived particles including jets and missing transverse momentum over a large range of $c\tau $. The search is optimized to constrain the decay of NLSP neutralino to a gravitino and either a Higgs boson or a Z boson, producing a final signal similar to the one in Fig.~\ref{fig:decay-chan} left. However, the more stringent limit over the signal cross-section obtained is 0.8 fb which is larger than the production cross-section corresponding to the lightest allowed gluino LSP.

\medskip
\noindent
\section{Strategy for the Scanning}
\label{strategy}

In this section, we describe the methodology that we have employed to search for points in our
parameter space that are compatible with the current experimental data on neutrino and Higgs physics, as well as ensuring that the gluino is the LSP.
In addition, we have demanded the compatibility with some flavour observables,
such as $B$ and $\mu$ decays.
To this end, we have performed scans on the parameter space of the model, with the input parameters optimally chosen.

\subsection{Experimental Constraints}
\label{sec:constr}

All experimental constraints (except the LHC searches, which are discussed in the previous section) are taken into account as follows:

\begin{itemize}

\item Neutrino observables\\
We have imposed the results for normal ordering from Ref.~\cite{Esteban:2018azc}, selecting points from the scan that lie within $\pm 3 \sigma$ of all neutrino observables. On the viable obtained points we have imposed the cosmological upper
bound on the sum of the masses of the light active neutrinos given
by $\sum m_{\nu_i} < 0.12$ eV~\cite{Aghanim:2018eyx}.

\item Higgs observables\\
The Higgs sector of the $\mn$ is extended with respect to the (N)MSSM.
For constraining the predictions in that sector of the model, we have interfaced 
{\tt HiggsTools} \cite{Bahl:2022igd} with {\tt Multinest}, using a 
conservative $\pm 3 \gev$ theoretical uncertainty on the SM-like Higgs boson in the $\mn$ as obtained with {\tt SPheno}. {\tt HiggsTools} is a unification of HiggsBounds-5~{\cite{Bechtle:2008jh,Bechtle:2011sb,Bechtle:2013wla,Bechtle:2015pma,Bechtle:2020pkv, Bahl:2021yhk}} and
HiggsSignals-2~{\cite{Bechtle:2013xfa,Bechtle:2020uwn}}. 
Our requirement is that the $p$-value reported by {\tt HiggsTools} be larger than 5\%, which is equivalent to impose $\chi^2 < 209$ for the 159 relevant degrees of freedom taken into account in our numerical calculation.

\item $B$~decays\\
$b \to s \gamma$ occurs in the SM at leading order through loop diagrams.
We have constrained the effects of new physics on the rate of this 
process using the average {experimental value of BR$(b \to s \gamma)$} $= (3.55 \pm 0.24) \times 10^{-4}$ provided in Ref.~\cite{Amhis:2012bh}. 
Similarly to the previous process, $B_s \to \mu^+\mu^-$ and  $B_d \to \mu^+\mu^-$ occur radiatively. We have used the combined results of LHCb and CMS~\cite{CMSandLHCbCollaborations:2013pla}, 
$ \text{BR} (B_s \to \mu^+ \mu^-) = (2.9 \pm 0.7) \times 10^{-9}$ and
$ \text{BR} (B_d \to \mu^+ \mu^-) = (3.6 \pm 1.6) \times 10^{-10}$. 
We put $\pm 3\sigma$ cuts from $b \to s \gamma$, $B_s \to \mu^+\mu^-$ and $B_d \to \mu^+\mu^-$, {as obtained with {\tt SPheno}}. We have also checked that the values obtained are compatible with the $\pm 3 \sigma$ of the results from the LHCb collaboration \cite{Santimaria:2021}, and with the combination of results on the rare $B_s^0 \to \mu^+ +\mu^-$ and $B^0 \to \mu^+ +\mu^-$ decays from the ATLAS, CMS, and LHCb \cite{CMS:2020rox}. 

\item $\mu \to e \gamma$ and $\mu \to e e e$\\
We have also included in our analysis the constraints {from 
BR$(\mu \to e\gamma) < 4.2\times 10^{-13}$~\cite{TheMEG:2016wtm}}
and BR$(\mu \to eee) < 1.0 \times 10^{-12}$~\cite{Bellgardt:1987du}, {as obtained with {\tt SPheno}}.

\item Chargino mass bound\\

Charginos have been searched at LEP with the result of a lower limit on the lightest chargino mass of 103.5 GeV in RPC MSSM, assuming universal gaugino and sfermion masses 
at the GUT scale and electron sneutrino mass larger than 300 GeV~\cite{wg1}. This limit is affected if the mass difference between chargino and neutralino is small, and the lower bound turns out to be in this case 92 GeV~\cite{wg2}. LHC limits  can be stronger but for very specific mass relations~\cite{ATLAS:2019lff,CMS:2018xqw,ATLAS:2017qwn,CMS:2018yan}.
Although in the $\mn$ there is RPV and therefore these constraints do not apply automatically, we usually choose in our analyses of the model the conservative limit of $m_{\widetilde \chi^\pm_1} > 92 \gev$.
However, since in this work we are analysing the gluino as the LSP, the chargino mass is always well above the mentioned bound.

\item Electroweak precision measurements\\
There have been recently several improvements in EW measurements such as $M_W$, $g-2$, $S, T, U$, etc. (see e.g. Refs.~\cite{Heinemeyer:2006px,Chakraborti:2022vds,Cho:2011rk}). Thus, the confrontation of the theory predictions and experimental results might be timely for SUSY models. However, we do not impose them in our scans, because in our framework electroweak precision measurements are not given significant contributions. This is because,
as we will see in Sec.~\ref{sec:results},
the SUSY mass spectrum turns out to be above 2.6 TeV, where the latter value is the lower bound that we obtain for the mass of the gluino LSP satisfying the LHC constraints. 
Furthermore, the masses of the neutralinos must be larger than the mass of the gluino, implying large $\mu$-term and Majorana masses, which in turn implies large masses for the Higgses (except the SM-like one), well above the gluino mass.

\end{itemize}


\subsection{Parameter Analysis}
\label{sec:parameter}

\noindent
We performed scans using the minimal possible set of parameters necessary to identify points with the gluino as the LSP. The entire mass spectrum was obtained using the full one-loop numerical computation implemented in \texttt{SPheno}. In what follows, we employed the following steps: \\

\noindent
{\bf Step 1}: The main goal of this step is to find benchmark points satisfying neutrino physics. For this, we started by optimizing the number of parameters of the scan by scanning over the VEVs $v_{1L}$ and $v_{2L}$ and taking the third left sneutrino VEV as $v_{3L} = 1.8\ v_{2L}$, a relation that is inspired by the specific solution for neutrino physics discussed in Ref.~\cite{Kpatcha:2019gmq}. Also, we scanned over $Y_{\nu_{1}}$, take $Y_{\nu_{2}} = 1.8\ Y_{\nu_{1}}$ and fix $Y_{\nu_{3}}$, which allows us to reproduce the normal ordering of neutrino masses with $Y_{\nu_3} < Y_{\nu_1} <  Y_{\nu_2}$. Finally, we scanned over the soft bino mass $M_1$, which is sufficient to reproduce neutrino data, and fix $M_2$.

Concerning Higgs physics, the soft stop masses $m_{\widetilde Q_{3L}}$ and $m_{\widetilde u_{3R}}$, and the trilinear parameter $T_{u_3}$ are very important for obtaining the correct mass for the SM-like Higgs, as discussed in Sec.~\ref{sec:model}. In addition, reproducing Higgs data requires suitable additional parameters $\lambda$, $\kappa$, $\tan\beta$, $v_R$, $T_\kappa$ and $T_\lambda$. Nevertheless, we fix $\kappa$, $T_\kappa$ and $v_R$, which basically control the right sneutrino sector, while suitably varying $\lambda$, $T_\lambda$, and $\tan\beta$.

Summarizing, we performed  two scans over the 9~parameters discussed above, as shown in Table~\ref{Scans-parameters}.
For Scan 1, the soft mass of the squark doublet is taken as $m_{\widetilde Q_{3L}} \in (200, 1200)$ GeV, whereas the soft mass of the right stop is fixed to a large value, $m_{\widetilde u_{3R}} = 2000$ GeV, as shown in 
Table~\ref{Fixed-parameters}.
Scan 2 is similar but with the replacement $m_{\widetilde Q_{3L}}\leftrightarrow m_{\widetilde u_{3R}}$.
In Table~\ref{Fixed-parameters}, we also show 
the remaining benchmark parameters, which are fixed to appropriate values. Note that at this step, the value of $M_3$ is also fixed according to Table~\ref{Fixed-parameters}.
\\

\begin{table}[t!]
\begin{center}
\begin{tabular}{|l|l|l|l|}
\hline
 \multicolumn{2}{|c|}{\bf Scan 1} &  \multicolumn{2}{|c|}{\bf Scan 2}\\ 
\hline
 \multicolumn{2}{|c|}{ $m_{\widetilde Q_{3L}} \in (200, 1200)$} &  \multicolumn{2}{|c|}{ $m_{\widetilde u_{3R}} \in (200, 1200)$}\\ 
\hline
\multicolumn{4}{|c|}{ $Y_{\nu_{1}} \in (10^{-8} , 10^{-6})$ }\\
\multicolumn{4}{|c|}{ $v_{1,2L} \in (10^{-5} , 10^{-3})$  }\\
\multicolumn{4}{|c|}{ $M_1 \in (1500 , 2500)$ }\\
\multicolumn{4}{|c|}{ $T_\lambda  \in  (0.5 , 2000)$ }\\
\multicolumn{4}{|c|}{ $\lambda \in (0.3 , 0.7)$ }  \\ 
\multicolumn{4}{|c|}{ $\tan\beta \in (1 , 20)$  }  \\ 
\multicolumn{4}{|c|}{ $-T_{u_3}\in (0 , 2000)$  }  \\ 
\hline
\end{tabular}
\end{center}
  \caption{Range of low-energy values of the input parameters that are varied in the two scans. 
  The VEVs $v_{1L}, v_{2L}$ and the soft SUSY-breaking parameter $m_{\widetilde Q_{3L}}$, $ m_{\widetilde u_{3R}}$, $T_{u_{3}}$, $M_1$ and $T_\lambda$ are given in GeV. Relations $v_{3L} = 1.8 \, v_{2L}$ and $Y_{\nu_{2}} = 1.8 \, Y_{\nu_{1}}$ are used.}
 \label{Scans-parameters}
\end{table} 

\begin{table}[t!]
\begin{center}
\begin{tabular}{|l|l|}
\hline
 \multicolumn{1}{|c|}{\bf Scan 1 }&\multicolumn{1}{c|}{ \bf Scan 2}\\
\cline{1-2}
   $m_{\widetilde u_{3R}} = 2000 $ &  \,\,\,\,\,\, $m_{\widetilde Q_{3L}}  = 2000$\\
\hline
\multicolumn{2}{|c|}{ $\kappa$ = 0.5} \\ 
\multicolumn{2}{|c|}{ $-T_{\kappa}$ = 300 }\\ 
\multicolumn{2}{|c|}{ $Y_{\nu_{3}}= 10^{-8}$ }\\ 
\multicolumn{2}{|c|}{ $ v_R$ = 3500} \\ 
\multicolumn{2}{|c|}{  $M_2$ = 2000, $M_3$ = 2700} \\ 
\multicolumn{2}{|c|}{$m_{\widetilde Q_{1,2\,L}}=m_{\widetilde u_{1,2\,R}} = m_{{\widetilde{d}}_{1,2,3R}} =m_{\widetilde e_{1,2,3\,R}}=$ 2000 } \\
\multicolumn{2}{|c|}{ $T_{u_{1,2}}=T_{d_{1,2}}=T_{e_{1,2}}=0$} \\
\multicolumn{2}{|c|}{$T_{d_{3}}=$ 100, $T_{e_{3}}=$ 40 }\\
\multicolumn{2}{|c|}{ $-T_{\nu_{1,2,3}}= 10^{-3}$}    \\ 
\cline{1-2}
\end{tabular}
\end{center}
  \caption{Low-energy values of the input parameters that are fixed in the two scans. The VEV $v_R$ and the soft trilinear parameters, soft wino and gluino masses and soft scalar masses are given in GeV.}
 \label{Fixed-parameters}
\end{table}

\noindent
{\bf Step 2}: 
Next, we took advantage of the fact that the parameters controlling neutrino and Higgs physics are essentially decoupled. Of the points derived in {\it Step 1}, we selected several starting points for the analysis with fixed values of the parameters that satisfy neutrino physics, namely $Y_{\nu_{1}}$, $v_{1L}$, $v_{2L}$, and $M_1$.For these points, we also took the values of $Y_{\nu_{2}}$ and $v_{3L}$ following Table~\ref{Scans-parameters}, and $Y_{\nu_{3}}$ and $M_2$ as provided in Table~\ref{Fixed-parameters}. 
%
Subsequently, we performed a scan over a subset of the parameters, which are relevant for Higgs physics. Specifically, we varied $m_{\widetilde Q_{3L}}$ ($m_{\widetilde u_{3R}}$) in the range  $900-2000$ GeV and $m_{\widetilde u_{3R}}$ ($m_{\widetilde Q_{3L}}$) in the range $2000-3000$ GeV. Additionally, we set the value of $T_{u_3}$ to be as low as $-3000$ GeV. As for the remaining parameters in Table~\ref{Scans-parameters}, namely $\lambda$, $T_\lambda$, and $\tan\beta$, we selected specific appropriate values for each of the starting points in the analysis.
%
Furthermore, in order to obtain the gluino as the LSP, we varied $M_3$ in the range of $500-2000$ GeV in contrast to {\it Step 1} where it was fixed to a concrete value (see Table~\ref{Fixed-parameters}), and set $T_\kappa = -800$ GeV to increase the mass of the pseudoscalar right sneutrinos, and $T_{\nu_{1,2,3}} = -10^{-2}$ GeV to increase the masses of the left sneutrinos and sleptons. Note that with the choice of ranges of different parameters, the stop left (right) is the NLSP in Scan 1 (2). \\

\noindent
{\bf Step 3}: 
As we will see in the next section, the comparison with LHC data of the results for the points derived from the previous steps, implies that all points found compatible with the experimental constraints of Sec.~\ref{sec:constr}, turn out to be excluded. Given this strong constraint, we considered the case of a gluino LSP with a larger mass, namely with soft mass in the range of $1900 - 3000$ GeV. The ranges of values of varied parameters, and the values of fixed parameters, are summarized in Tables~\ref{S3-Scans-priors-parameters} and~\ref{S3-Scans-fixed-parameters}, respectively.

 \begin{table}[t!]
\begin{center}
\begin{tabular}{|l|l|l|l|}
\hline
\multicolumn{4}{|c|}{ \bf Scan 3 } \\
\hline
   \multicolumn{4}{|c|}{  $ M_3 \in (1900, 3000) $ } \\
   \multicolumn{4}{|c|}{  $ m_{\widetilde u_3} \in (1900, 3000) $ } \\
   \multicolumn{4}{|c|}{  $ m_{\widetilde Q_3} \in (1900, 3000) $ } \\
   \multicolumn{4}{|c|}{  $ - T_{u_{3}} \in (0, 5000) $ } \\
   \multicolumn{4}{|c|}{  $ T_\lambda \in (0, 4000) $ } \\
   \multicolumn{4}{|c|}{  $\tan\beta \in (1, 10)$  }  \\  
  \hline
\end{tabular}
\end{center}
  \caption{Range of low-energy values of the input parameters that are varied in Scan 3. 
  The soft SUSY-breaking parameters $m_{\widetilde Q_{3L}}$, $ m_{\widetilde u_{3R}}$, $T_{u_{3}}, T_\lambda$, and $M_3$ are given in GeV.}
  \label{S3-Scans-priors-parameters}
\end{table} 

\begin{table}[t!]
\begin{center}
\begin{tabular}{|c|c|c|c|}
\hline
\multicolumn{4}{|c|}{ \bf Scan 3 } \\
\hline 
\multicolumn{4}{|c|}{ $\lambda$ = 0.33,  $\kappa$ = 0.6, $v_R$ = 4500 } \\ 
\multicolumn{4}{|c|}{ $Y_{\nu_{1}}= 8.82 \times 10^{-7}$,   $Y_{\nu_{2}}= 1.62 \times 10^{-6}$,  $Y_{\nu_{3}}= 4.42 \times 10^{-7}$ }\\ 
\multicolumn{4}{|c|}{  $T_{u_{1,2}}=T_{d_{1,2}}=T_{e_{1,2}}=0$} \\
\multicolumn{4}{|c|}{  $T_{d_{3}}=$ 100, $T_{e_{3}}=$ 40 } \\
\multicolumn{4}{|c|}{ $-T_{\nu_{1,2,3}}= 0.1$}    \\ 
\multicolumn{4}{|c|}{ $-T_{\kappa} = 1200$} \\  
\multicolumn{4}{|c|}{$m_{\widetilde Q_{1,2L}}=m_{\widetilde u_{1,2R}}=m_{\widetilde d_{1,2,3R}}=m_{\widetilde e_{1,2,3R}}=$ 4500 GeV } \\
\multicolumn{4}{|c|}{  $M_1$ = 3480 GeV, $ M_2$ = 3620 GeV} \\  
\hline
\end{tabular}
\end{center}
  \caption{Low-energy values of the input parameters that are fixed in Scan 3. The VEV $v_R$ and the soft trilinear parameters, and soft SUSY-breaking parameters are given in GeV.}
\label{S3-Scans-fixed-parameters}
\end{table}

\section{Results}
\label{sec:results}
Following the methods described in the previous sections, in order
to find regions consistent with experimental observations we performed scans of the parameter space, and our results are presented here.
To carry this analysis out, we selected first points from the scans that lie within $\pm 3\sigma$ of all neutrino physics observables \cite{Esteban:2018azc}.
Second, we put $\pm 3\sigma$ cuts from $b \to s \gamma$, $B_s \to \mu^+\mu^-$ and $B_d \to \mu^+\mu^-$ 
and require the points to satisfy also the upper limits of $\mu \to e \gamma$ and $\mu \to eee$. 
In the third step, we imposed that Higgs physics is realized.}
In particular, we require that the p-value reported by {\tt HiggsTools} be larger
than 5\%. 

\begin{figure}[t!]
\centering
\includegraphics[width=\linewidth]{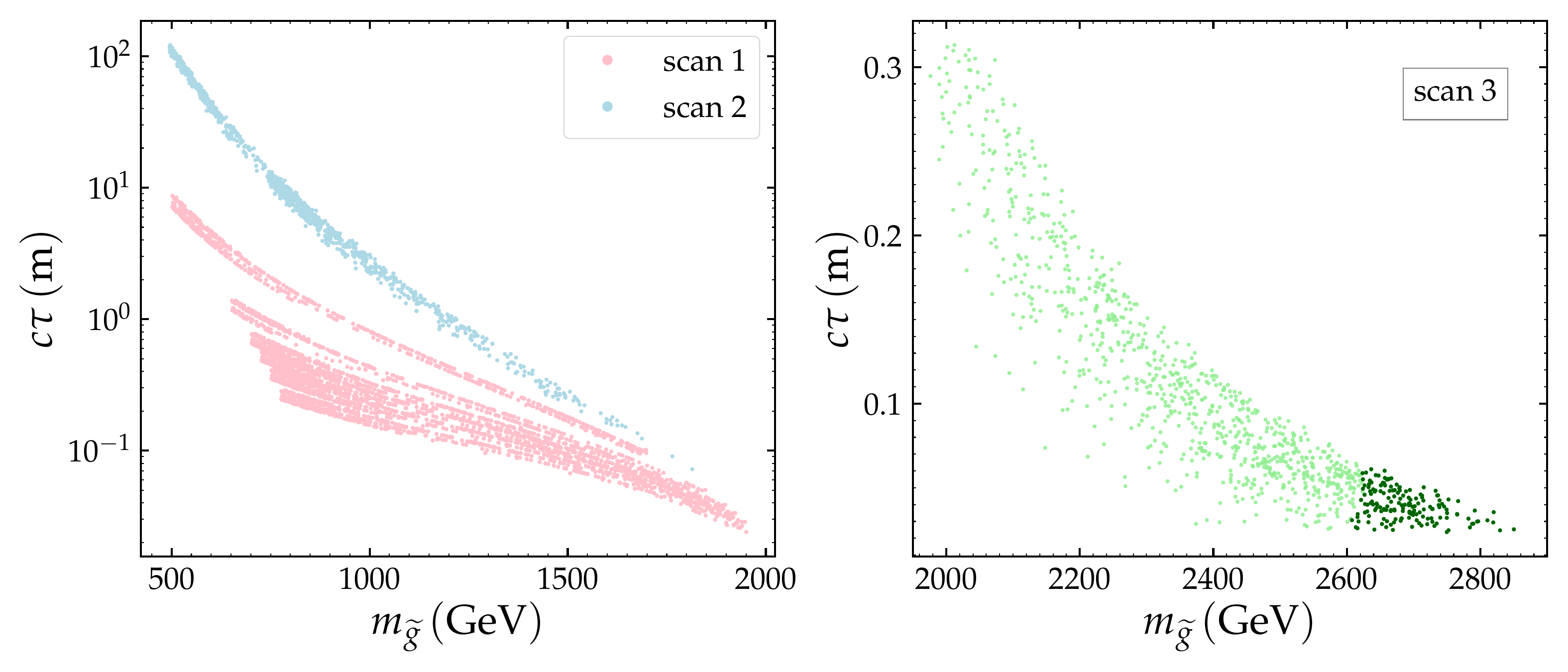}
  \caption{Gluino proper decay length $c\tau$ versus the gluino mass
  $m_{\widetilde g}$ for (left plot) Scans 1, 2 shown in Tables~\ref{Scans-parameters} and~\ref{Fixed-parameters}, and for (right plot) Scan 3 shown in Tables~\ref{S3-Scans-priors-parameters} and~\ref{S3-Scans-fixed-parameters}. 
  All points fulfil the experimental constraints of Sec.~\ref{sec:constr}.
  Cyan, pink and light green points do not fulfil the LHC constraints, while dark green ones do.}
\label{LHC-ctau-vs-mGlu}
\end{figure}

In Fig.~\ref{LHC-ctau-vs-mGlu}, we show the decay length of a gluino LSP for the three scans conducted. The left plot pertains to Scans 1 and 2, while the right plot shows the results from Scan~3. As can be seen, all points analysed satisfy the experimental constraints discussed in Sec.~\ref{sec:constr}. 
However, all (pink) points in Scan 1, as well as all (cyan) points in Scan 2, corresponding to gluino masses $m_{\tilde g} \lsim 2000$ GeV, turn out to be excluded because of the LHC constraints discussed in Sec.~\ref{sec:lhc}. 
From Case (i), all points with $c\tau> 3$m and $m_{\tilde{g}}<2000$ GeV are excluded because of large ionization energy loss. The rest of the points are excluded
by the searches for displaced vertices and non-prompt jets corresponding to Cases (ii) to (v).

These results motivated Scan 3, where higher values of $M_3 \lesssim 3000$ GeV were investigated. As can be seen in the right panel, the largest mass of the gluino as the LSP is~$\sim 2850$ GeV, below the upper limit of $M_3 = 3000$ GeV for this scan, since above this value the gluino is no longer the LSP.
Constraints from Cases (ii) to (v) exclude the light green points with 
gluino masses up to $m_{\tilde{g}} \sim 2607$ GeV. 
Thus, our findings reveal that some regions of the parameter space for the gluino LSP are still compatible with LHC searches, as represented by the dark green points.
Note that, the minimum decay length observed is approximately $c\tau \sim 23$ mm for $m_{\tilde{g}} \sim 2850$ GeV, indicating that the prompt decay searches of Case (vi) in Sec.~\ref{sec:lhc} are not applicable to our analysis.

In general, the problem to further constrain the lower limit on the gluino mass
is that for large masses, the number of predicted events is very scarce ($\sim 5$ events at 137 fb$^{-1}$ before any selection), and there are simply not enough events to push the limit further, regardless of the analysis. The maximum reachable limit with an ideal search at 13 TeV and 137 fb$^{-1}$ would be around 2750 GeV actually, where one expects $\sim 3$ events before selections.

On the other hand, according to Ref.~\cite{ATLAS:2018yii}, at the HL-LHC, the discovery reach and expected limit of displaced vertex searches in terms of the gluino mass are expected to be improved by $\sim 500$~GeV and $\sim 1000$~GeV, respectively, compared to the current limit~\cite{ATLAS:2017tny}. Therefore, the remaining points in Fig.~\ref{LHC-ctau-vs-mGlu} can fully be tested in the HL-LHC experiment.

\begin{figure}[t!]
\centering
\includegraphics[width=\linewidth]{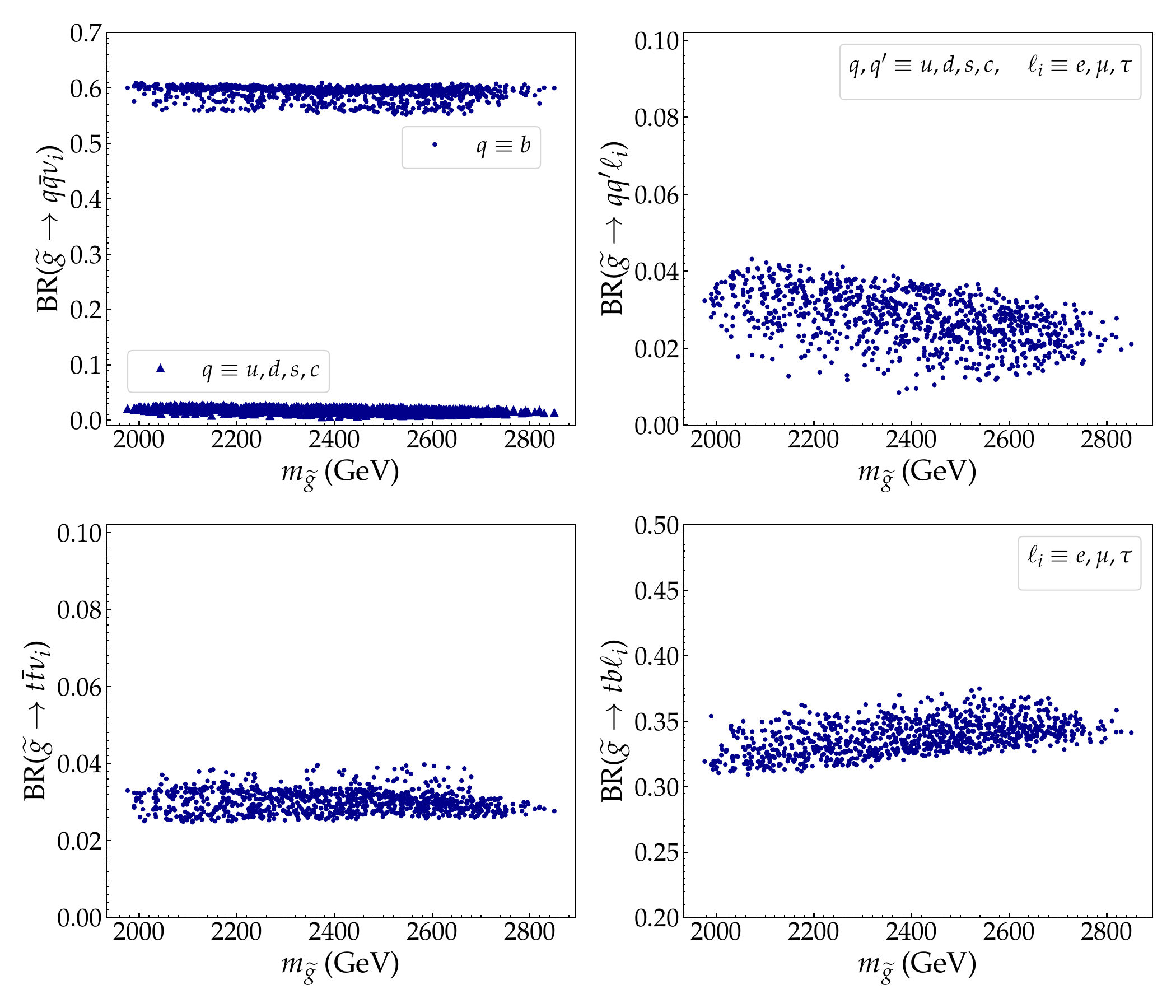}
  \caption{
  For the points presented in the right panel of Fig.~\ref{LHC-ctau-vs-mGlu}, we show the
  sum of branching ratios of gluino LSP decaying to (left plots) quark-quark-neutrino and (right plots) quark-quark-lepton, versus the gluino mass $m_{\widetilde g}$. The latter include both the decays into leptons and anti-leptons. 
  }
\label{LHC-BRs-MGlu}
\end{figure}

Summarizing, we found that the upper limit for the decay length of the allowed dark green points in the right panel of Fig.~\ref{LHC-ctau-vs-mGlu}
is $c\tau \simeq 0.061$m, which corresponds to a lower limit on the gluino mass of $m_{\tilde{g}} \simeq 2607$ GeV.

In Fig.~\ref{LHC-BRs-MGlu}, the relevant branching ratios (BRs) corresponding to the light and dark green points from the right plot of Fig.~\ref{LHC-ctau-vs-mGlu} (Scan 3) are presented.  As anticipated in Sec.~\ref{subsec:Decaymodes}, the dominant decay channels involve third-generation quarks, irrespective of whether neutrinos or leptons are in the final states. In particular,
the BR($\widetilde{g} \to q \bar{q}^{\prime} \ell$) constitutes approximately 10\% of the BR($\widetilde{g} \to t \bar{b} \ell$).
Besides, the 
BR($\widetilde{g} \to q \bar{q} \nu$), where $q$ includes the top quark, constitutes approximately 3\% of BR($\widetilde{g} \to b \bar{b} \nu$).
Concerning the latter BRs,
it is noteworthy that the BR($\widetilde{g} \to t \bar{t} \nu$) is significantly smaller, about 5\% of BR($\widetilde{g} \to b \bar{b} \nu$). This difference arises because, unlike Eq.~(\ref{gamma-ttnu}), the dominant contribution in Eq.~(\ref{gamma-qqnu}) is proportional to $Y_b$. Consequently, for first- and second-generation quarks, where $Y_q$ is negligible, the BR of $\widetilde{g} \to q \bar{q} \nu$ (represented as triangles in the upper plot of the left panel) 
and $\widetilde{g} \to t \bar{t} \nu$ are expected to be of similar magnitude.

Additionally, it is worth noting that within the parameter space explored in Scan 3, the BRs exhibit minimal variations. The small differences observed are attributed to variations in the mixing elements between neutrinos and leptons with bino/winos, corresponding to different values of the left neutrino VEVs, $v_{iL}$. Also, let us remark that the one-loop decay to gluon and neutrino is significantly suppressed and can be considered negligible.

Finally, let us comment that the use of other type of solutions for neutrino physics different from the one presented in Eq.~(\ref{neutrinomassess}), would not modify the results obtained. This can be understood from the summation over leptons present in Eqs.~(\ref{--sneutrino-decay-width-2nus2}-\ref{gamma-qqnu}), since for the most restrictive searches, for instance ~\cite{CMS:2019qjk, ATLAS:2017tny, CMS:2020iwv},
the results are independent of the lepton family or integrate over it. In the particular case of the search~\cite{ATLAS:2020xyo}, where there are muons in the final state, the above conclusion remains valid because the bounds are far from restrictive.


\section {Conclusions}
\label{sec:conclusion}
We analysed the signals expected at the LHC for a gluino LSP in the framework of the $\mn$, imposing on the parameter space the experimental constraints on neutrino and Higgs physics, as well as flavour observables such as $B$ and $\mu$ decays. 
The gluino has three relevant decay modes.
On the one hand, it has two three-body decays to two quarks and a lepton or a neutrino, mediated by virtual squarks.  
The dominant channels correspond to the third family of quarks.
On the other hand, the gluino can decay to gluon and neutrino through a virtual quark-squark loop. 
However, given the large values of the gluino masses analysed, this channel is suppressed compared to the three-body decays. We studied these channels and the corresponding decay lengths for three relevant scans of the parameter space. For two of the scans, we analysed the range for the gluino mass parameter of $M_3 = 500 - 2000$ GeV. We obtained that all points are excluded by the LHC searches of Ref.~\cite{ATLAS:2022pib}, where a large ionization energy loss is considered.
In the third scan, we analysed the wider range of $M_3 = 1900 - 3000$ GeV, obtaining an allowed region. Specifically, searches for displaced vertices and non-prompt jets~\cite{CMS:2019qjk, ATLAS:2017tny, CMS:2020iwv, ATLAS:2020xyo, ATLAS:2020wjh}
exclude gluino masses up to 2607 GeV,
as shown in Fig.~\ref{LHC-ctau-vs-mGlu}. Our results also imply an upper limit for the gluino decay length of 6.1 cm.


\begin{acknowledgments}
The work of PK and DL was supported by the Argentinian CONICET, and they also acknowledge the support through {PICT~2020-02181}
The work of EK was supported by the grant "Margarita Salas" for the training of young doctors (CA1/RSUE/2021-00899), co-financed by the Ministry of Universities, the Recovery, Transformation and Resilience Plan, and the Autonomous University of Madrid. 
{The work of IL was funded by the Norwegian Financial Mechanism 2014-2021, grant DEC-2019/34/H/ST2/00707.}
The research of CM was partially supported by the AEI through the grants IFT Centro de
Excelencia Severo Ochoa No CEX2020-001007-S and PID2021-125331NB-I00, funded by
MCIN/AEI/10.13039/501100011033.
The work of NN was supported by JSPS KAKENHI Grant Number 21K13916. 

\end{acknowledgments}

\appendix
\numberwithin{equation}{section}
\numberwithin{figure}{section}
\numberwithin{table}{section}


\section{Validation of the Search for Displaced Leptons} 
\label{appendix}

We tested the implementation of the search using Monte Carlo samples generated for the process $pp\to \widetilde{l}^*\widetilde{l}$ up to two additional partons in the hard event, where the slepton is either a selectron or smuon and further decays into a neutralino and a electron or muon, respectively. Samples are generated for masses between 50 and 800 GeV and considering proper decay lengths between 3 and 3000 mm. The tree-level sample events are generated with MadGraph5\_aMC@NLO-v3.4.2~\cite{Alwall:2014hca,Alwall:2007fs,Alwall:2008qv} at leading order (LO).  Simulated signal events were passed to 
Pythia-8.306~\cite{Sjostrand:2014zea} for parton showering (PS) and hadronization. Jet matching and merging to parton-shower calculations is accomplished by the MLM algorithm~\cite{Mangano:2006rw}.
Slepton pair-production nominal cross-sections are derived at NLO using Resummino 3.1.2~\cite{Debove:2011xj,Debove:2009ia,Debove:2010kf,Fuks:2013vua,Fuks:2012qx,Fiaschi:2018xdm,Fiaschi:2020udf}.\\

\noindent
{\bf Event selection}: 
The event selection utilizes the data provided by the ATLAS collaboration in the auxiliary material. First, it applies a preselection step for truth-level particles, based on a parametrization of the reconstruction and identification efficiency of leptons reported by ATLAS, and requires for the surviving preselected leptons $p_T^l>65$ GeV and $3\ mm<|d_0|<300\ mm$. Events are then required to have exactly two reconstructed leptons. Then, it applies a parameterized event-level selection based on the acceptance maps reported by ATLAS, as a function of the generator-level $p_T$ of the leading and sub-leading lepton. Additional isolation requirements are placed on the leptons by requiring that the sum of $p_T$ of all
tracks that fall within some $\Delta R$ of the lepton track are smaller than a fraction f of the lepton-$p_T$ and by requiring that the sum energy deposits near the lepton in the calorimeters must be less than a fraction of the energy of the lepton. Here are the requirements in terms of ($\Delta R$, f) for the different cases: 
\begin{itemize}
    \item Electrons track: $(0.2, 0.06)$
    \item Electrons calorimeter: $(0.2, 0.06)$
    \item Muons track: $(0.3, 0.04)$
    \item Muons calorimeter: $(0.2, 0.15)$
\end{itemize}

Finally, signal leptons are required to be separated by $\Delta R>0.2$.

Numbers in Tables~\ref{validation_sel_table},~\ref{validation_smu_table} and~\ref{validation_stau_table}, and Fig.~\ref{validation} show a comparison between the ATLAS results and the recast version of the search. The agreement is generally good but shows a moderately smaller values for lifetimes $0.1-1$ ns, coinciding with the largest sensitivity of the search, and shows a larger deviation of the ATLAS sensitivity for values of the lifetime between  $0.01-0.1$ ns.

\begin{table}[H]
\begin{center}
\begin{tabular}{|l|l|l|l|}
\hline
 \multicolumn{1}{|c|}{\bf Selection} &  \multicolumn{3}{c|}{\bf $\widetilde{e}$ (mass [GeV], lifetime [ns])}\\ 
\multicolumn{1}{|c|}{}&  \multicolumn{1}{c}{(100, 0.01)}&  \multicolumn{1}{c}{(300, 1)}&  \multicolumn{1}{c|}{(500, 0.1)} \\
\hline
initial number of events ($\mathcal{L}\times \sigma$)&50830&870&93.6\\
final selection (ATLAS)&77.1&52.2&17.7\\
final selection (CheckMATE)&35.0&47.4&10.1\\
difference [$\%$]&-54.6&-9.3&-43.1\\
 \hline
\end{tabular}
\end{center}
  \caption{Comparison of the number of expected events in the $ee$ signal region for the process $pp\to \widetilde{e}^*\widetilde{e}$.}
 \label{validation_sel_table}
\end{table}

\begin{table}[H]
\begin{center}
\begin{tabular}{|l|l|l|l|}
\hline
 \multicolumn{1}{|c|}{\bf Selection} &  \multicolumn{3}{c|}{\bf $\widetilde{\mu}$ (mass [GeV], lifetime [ns])}\\ 
\multicolumn{1}{|c|}{}&  \multicolumn{1}{c}{(100, 0.01)}&  \multicolumn{1}{c}{(300, 1)}&  \multicolumn{1}{c|}{(500, 0.1)} \\
\hline
initial number of events ($\mathcal{L}\times \sigma$)&50830&870&93.6\\
final selection (ATLAS)&25.0&49.9&13.6\\
final selection (CheckMATE)&25.33&64.6&10.15\\
difference [$\%$]&1.3&29.5&-25.4\\
 \hline
\end{tabular}
\end{center}
  \caption{Comparison of the number of expected events in the $\mu\mu$ signal region for the process $pp\to \widetilde{\mu}^*\widetilde{\mu}$.}
 \label{validation_smu_table}
\end{table} 

\begin{table}[H]
\begin{center}
\begin{tabular}{|l|l|l|}
\hline
 \multicolumn{1}{|c|}{\bf Selection} &  \multicolumn{2}{c|}{\bf $\widetilde{\tau}$ (mass [GeV], lifetime [ns])}\\ 
\multicolumn{1}{|c|}{}&  \multicolumn{1}{c}{(200, 0.1)}&  \multicolumn{1}{c|}{(300, 0.1)} \\
\hline
initial number of events ($\mathcal{L}\times \sigma$)&4210&870\\
final selection (ATLAS)&3.45&1.26\\
final selection (CheckMATE)&2.44&0.55 \\
difference [$\%$]&-41&-56 \\
 \hline
\end{tabular}
\end{center}
  \caption{Comparison of the number of expected events in the $e \mu$ signal region for the process $pp\to \widetilde{\tau}^*\widetilde{\tau}$.}
 \label{validation_stau_table}
\end{table} 

\begin{figure}[H]
\centering
\includegraphics[width=0.47\linewidth]{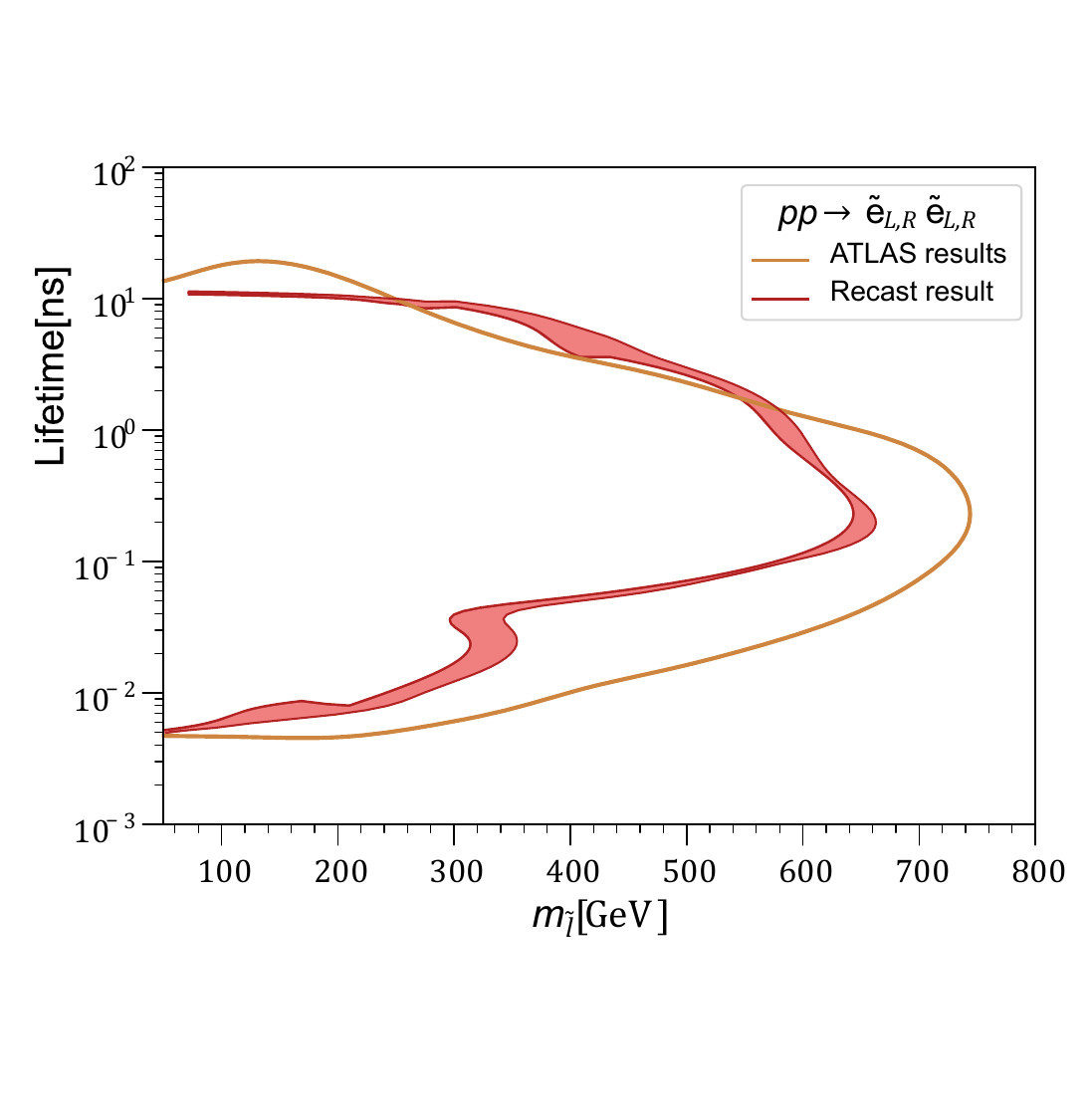}
\includegraphics[width=0.47\linewidth]{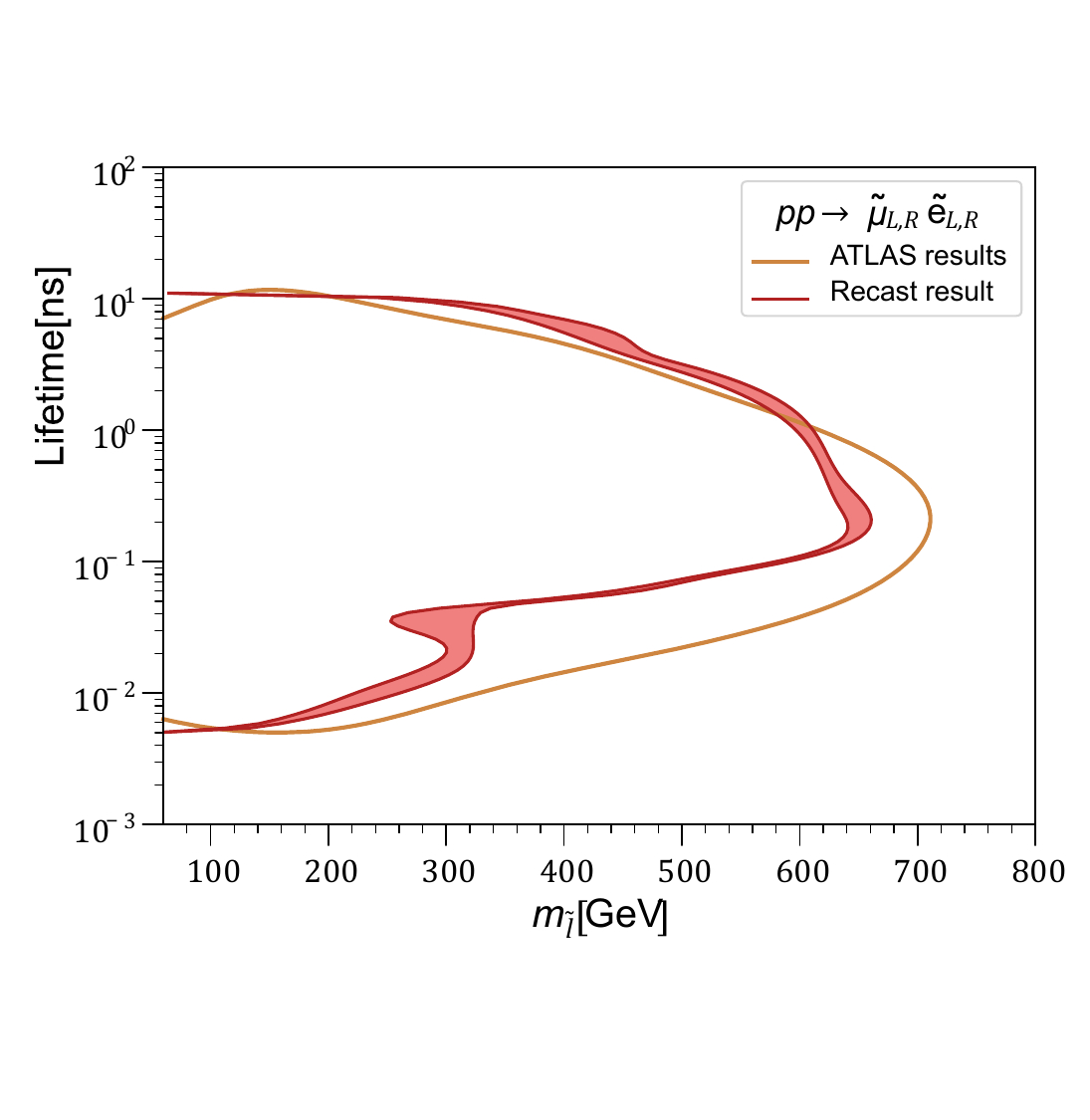}
  \caption{Comparison of the observed exclusion contours for selectron (smuon) NLSP on the left (right). The orange line shows the ATLAS results and the red line the limit obtained with CheckMATE.
  }
\label{validation}
\end{figure}

 \section{Gluino LSP Relevant Interactions}
 \label{appendixb}

In this Appendix, we write the relevant interactions for our computation of the gluino LSP decays, 
following {\tt SARAH} notation~\cite{Staub:2013tta}.
In particular, 
now
$a,b=1,2,3$ are family indexes, $i,j,k$ are the indexes for the physical states, and $\alpha, \beta,\gamma=1,2,3$ are $SU(3)_C$ indexes ($\alpha=1, ...,8$ when gluinos contribute to the interactions as in (\ref{gluinodown}) and (\ref{gluinoup}) below, where $\lambda^\alpha$ are the Gell-Mann matrices 
).
The matrices $Z^D, Z^U, U^d_{L,R}, U^u_{L.R}, U^e_{L,R}$ and $U^V$ diagonalize the mass matrices of down squarks, up squarks, down quarks, up quarks, charged fermions (leptons, gauginos and Higgsinos) and neutral fermions (LH and RH neutrinos, gauginos and Higgsinos), respectively.
More details about these matrices can be found in Appendix B of Ref.~\cite{Ghosh:2017yeh}.
Taking all this into account, in the basis of 4--component spinors with the projectors 
$P_{L,R}=(1\mp\gamma_5)/2$, the interactions for the mass eigenstates are as follows.

\subsection{Gluino - down quark - down squark Interaction}

\begin{center} 
\begin{fmffile}{Figures/Diagrams/FeynDia110} 
\fmfframe(20,20)(20,20){ 
\begin{fmfgraph*}(75,75) 
\fmfleft{l1}
\fmfright{r1,r2}
\fmf{plain}{l1,v1}
\fmf{fermion}{r1,v1}
\fmf{scalar}{v1,r2}
\fmflabel{$\tilde{g}_{{\alpha}}$}{l1}
\fmflabel{$d_{{j \beta}}$}{r1}
\fmflabel{$\tilde{d}^*_{{k \gamma}}$}{r2}
\end{fmfgraph*}} 
\end{fmffile} 
\end{center}  
\begin{align} 
 \nonumber &-i \frac{1}{\sqrt{2}} g_3
 \lambda^{\alpha}_{\gamma,\beta} \sum_{a=1}^{3}U^{d,*}_{L,{j a}} Z_{{k a}}^{D}  \Big(\frac{1-\gamma_5}{2}\Big)\\ 
  & + \,i \frac{1}{\sqrt{2}} g_3
  \lambda^{\alpha}_{\gamma,\beta} \sum_{a=1}^{3}Z_{{k 3 + a}}^{D} U_{R,{j a}}^{d}  \Big(\frac{1+\gamma_5}{2}\Big)
  \label{gluinodown}\end{align}

\subsection{Gluino - up quark - up squark Interaction}
 
\begin{center} 
\begin{fmffile}{Figures/Diagrams/FeynDia114} 
\fmfframe(20,20)(20,20){ 
\begin{fmfgraph*}(75,75) 
\fmfleft{l1}
\fmfright{r1,r2}
\fmf{plain}{l1,v1}
\fmf{fermion}{r1,v1}
\fmf{scalar}{v1,r2}
\fmflabel{$\tilde{g}_{{\alpha}}$}{l1}
\fmflabel{$u_{{j \beta}}$}{r1}
\fmflabel{$\tilde{u}^*_{{k \gamma}}$}{r2}
\end{fmfgraph*}} 
\end{fmffile} 
\end{center}  
\begin{align} 
 \nonumber &-i \frac{1}{\sqrt{2}} g_3 
 \lambda^{\alpha}_{\gamma,\beta} \sum_{a=1}^{3}U^{u,*}_{L,{j a}} Z_{{k a}}^{U}  \Big(\frac{1-\gamma_5}{2}\Big)\\ 
  & + \,i \frac{1}{\sqrt{2}} g_3 
  \lambda^{\alpha}_{\gamma,\beta} \sum_{a=1}^{3}Z_{{k 3 + a}}^{U} U_{R,{j a}}^{u}  \Big(\frac{1+\gamma_5}{2}\Big)
\label{gluinoup}\end{align}

\subsection{Up squark - down quark - lepton Interaction}
\label{SubSection:Coupling-Huu}

\begin{center} 
\begin{fmffile}{Figures/Diagrams/FeynDia103} 
\fmfframe(20,20)(20,20){ 
\begin{fmfgraph*}(75,75) 
\fmfleft{l1}
\fmfright{r1,r2}
\fmf{fermion}{v1,l1}
\fmf{fermion}{r1,v1}
\fmf{scalar}{r2,v1}
\fmflabel{$\bar{d}_{{i \alpha}}$}{l1}
\fmflabel{$e_{{j}}$}{r1}
\fmflabel{$\tilde{u}_{{k \gamma}}$}{r2}
\end{fmfgraph*}} 
\end{fmffile} 
\end{center}  
\begin{eqnarray}
 &&i \delta_{\alpha \gamma} U^{e,*}_{R,{j 5}}  \sum_{b=1}^{3}Z^{U,*}_{k b} \sum_{a=1}^{3}U^{d,*}_{R,{i a}} Y_{d,{a b}}\
 P_L
 \nonumber\\ 
  && \,-i \delta_{\alpha \gamma} \Big(g \sum_{a=1}^{3}Z^{U,*}_{k a} U_{L,{i a}}^{d}  U_{L,{j 4}}^{e}  - \sum_{b=1}^{3}\sum_{a=1}^{3}Y^*_{u,{a b}} Z^{U,*}_{k 3 + a}  U_{L,{i b}}^{d}  U_{L,{j 5}}^{e} \Big)
  P_R.
  \label{leptons}
\end{eqnarray}

\subsection{Down squark - up quark - lepton Interaction }

\begin{center} 
\begin{fmffile}{Figures/Diagrams/FeynDia119} 
\fmfframe(20,20)(20,20){ 
\begin{fmfgraph*}(75,75) 
\fmfleft{l1}
\fmfright{r1,r2}
\fmf{fermion}{v1,l1}
\fmf{fermion}{v1,r1}
\fmf{scalar}{r2,v1}
\fmflabel{$\bar{e}_{{i}}$}{l1}
\fmflabel{$\bar{u}_{{j \beta}}$}{r1}
\fmflabel{$\tilde{d}_{{k \gamma}}$}{r2}
\end{fmfgraph*}} 
\end{fmffile} 
\end{center} 

\vspace*{-0.5cm}

\begin{align} 
\nonumber  & i \delta_{\beta \gamma}\Big(U^{e,*}_{L,{i 5}}  \sum_{b=1}^{3}Z^{D,*}_{k b} \sum_{a=1}^{3}U^{u,*}_{R,{j a}} Y_{u,{a b}}\Big) P_L\\ 
  &  -\,i \delta_{\beta \gamma} \Big(g\sum_{a=1}^{3}Z^{D,*}_{k a} U_{L,{j a}}^{u}  U_{R,{i 4}}^{e}  - \sum_{b=1}^{3}\sum_{a=1}^{3}Y^*_{d,{a b}} Z^{D,*}_{k 3 + a}  U_{L,{j b}}^{u}  U_{R,{i 5}}^{e} \Big)P_R.
\label{eq:appen-ql}
  \end{align}

\subsection{Up squark - up quark - neutrino Interaction}

\begin{center} 
\begin{fmffile}{Figures/Diagrams/FeynDia109} 
\fmfframe(20,20)(20,20){ 
\begin{fmfgraph*}(75,75) 
\fmfleft{l1}
\fmfright{r1,r2}
\fmf{fermion}{v1,l1}
\fmf{plain}{r1,v1}
\fmf{scalar}{r2,v1}
\fmflabel{$\bar{u}_{{i \alpha}}$}{l1}
\fmflabel{$\nu_{{j}}$}{r1}
\fmflabel{$\tilde{u}_{{k \gamma}}$}{r2}
\end{fmfgraph*}} 
\end{fmffile} 
\end{center}  

\begin{eqnarray}
  &&\frac{i}{3} \delta_{\alpha \gamma} \Big(2 \sqrt{2} g' U^{V,*}_{j 4} \sum_{a=1}^{3}Z^{U,*}_{k 3 + a} U^{u,*}_{R,{i a}}   -3 U^{V,*}_{j 7} \sum_{b=1}^{3}Z^{U,*}_{k b} \sum_{a=1}^{3}U^{u,*}_{R,{i a}} Y_{u,{a b}}   \Big)P_L \nonumber \\  
   &&-\frac{i}{6} \delta_{\alpha \gamma} \Big[6 \sum_{b=1}^{3}\sum_{a=1}^{3}Y^*_{u,{a b}} Z^{U,*}_{k 3 + a}  U_{L,{i b}}^{u}  U_{{j 7}}^{V}  + \sqrt{2} \sum_{a=1}^{3}Z^{U,*}_{k a} U_{L,{i a}}^{u}  \Big(3 g U_{{j 5}}^{V}  + g' U_{{j 4}}^{V} \Big)\Big]P_R.\nonumber\\
  \label{neutrinos up}
  \end{eqnarray}

\subsection{Down squark - down quark - neutrino Interaction }
\begin{center} 
\begin{fmffile}{Figures/Diagrams/FeynDia108} 
\fmfframe(20,20)(20,20){ 
\begin{fmfgraph*}(75,75) 
\fmfleft{l1}
\fmfright{r1,r2}
\fmf{fermion}{v1,l1}
\fmf{fermion}{r1,v1}
\fmf{scalar}{r2,v1}
\fmflabel{$\bar{d}_{{i \alpha}}$}{l1}
\fmflabel{$\nu_{{j}}$}{r1}
\fmflabel{$\tilde{d}_{{k \gamma}}$}{r2}
\end{fmfgraph*}} 
\end{fmffile} 
\end{center} 

\vspace*{-0.5cm}
  \begin{align} 
\nonumber   &-\frac{i}{3} \delta_{\alpha \gamma} \Big(\sqrt{2} g' U^{V,*}_{j 4} \sum_{a=1}^{3}Z^{D,*}_{k 3 + a} U^{d,*}_{R,{i a}} + 3 U^{V,*}_{j 6} \sum_{b=1}^{3}Z^{D,*}_{k b} \sum_{a=1}^{3}U^{d,*}_{R,{i a}} Y_{d,{a b}}   \Big)P_L\\ 
    &  \,-\frac{i}{6} \delta_{\alpha \gamma} \Big(6 \sum_{b=1}^{3}\sum_{a=1}^{3}Y^*_{d,{a b}} Z^{D,*}_{k 3 + a}  U_{L,{i b}}^{d}  U_{{j 6}}^{V}  + \sqrt{2} \sum_{a=1}^{3}Z^{D,*}_{k a} U_{L,{i a}}^{d}  \Big(-3g U_{{j 5}}^{V}  + g' U_{{j 4}}^{V} \Big)\Big)P_R.
  \label{neutrinos down}
  \end{align} 

\clearpage


\bibliographystyle{utphys}
\bibliography{gluino-lsp_munuSSM}

\end{document}

%% file: defsV2.tex
\usepackage[T1]{fontenc}
\usepackage{epsfig}
\usepackage{latexsym}
\usepackage{graphicx}
\usepackage{amsmath}
\usepackage{amsfonts}   
\usepackage{amssymb}    
\usepackage{float}
\usepackage{bm}
\usepackage{url}
\usepackage{hyperref} 
\usepackage[nodisplayskipstretch]{setspace}
\usepackage{url}
\def\lsim{\raise0.3ex\hbox{$\;<$\kern-0.75em\raise-1.1ex\hbox{$\sim\;$}}}
\def\gsim{\raise0.3ex\hbox{$\;>$\kern-0.75em\raise-1.1ex\hbox{$\sim\;$}}}

\def    \beq            {\begin{equation}}
\def    \eeq            {\end{equation}}
\def    \bea           {\begin{eqnarray}}
\def    \eea           {\end{eqnarray}}

\def \mn{\mu\nu{\rm SSM}}

\def\g2{{\rm GeV}^2}

\def\sw2{sin^2 \theta_w}

\def\a^tau{\alpha_{\tau}}

\def\beq{\begin{equation}}
\def\eeq{\end{equation}}
\def\beqa{\begin{eqnarray}}
\def\eeqa{\end{eqnarray}}

\newcommand{\tev}{\;\textrm{TeV}}
\newcommand{\gev}{\;\textrm{GeV}}

\newcommand{\newc}{\newcommand}
\newc\BR{BR}
\newc{\akappa}{A_{\kappa} }
\newc\deltagmtwo{\delta (g-2)_{\mu}} 
\newc\deltaamu{\Delta a_{\mu}}

\def\anti{\overline}

\def\la{\lambda}

\def\ka{\kappa}

\newc{\haa}{BR\(h_1\to a_1 a_1\)}
\newc{\abb}{BR\(a_1\to b\anti{b}\)}
\newc{\hbb}{BR\(h_1\to b\anti{b}\)}
\newc{\abund}{\Omega h^2}
\newc\bsgamma{b\rightarrow s \gamma }
\newc\bxsgamma{\overline{B}\rightarrow X_{s}\gamma}
\newc\brbsgamma{\BR(\overline{B}\rightarrow X_s\gamma)}

\newcommand\ReDiag{\mathop{%
  \raise .5pt\hbox{[}%
  \widetilde{\mathrm{Re}}%
  \raise .5pt\hbox{]}}}
\newcommand\ReOffDiag{\mathop{%
  \raise .5pt\hbox{$\llbracket$}%
  \widetilde{\mathrm{Re}}%
  \raise .5pt\hbox{$\rrbracket$}}}

\def\order#1{\ensuremath{{\cal O}(#1)}}

\def\la{\lambda}

\def\gmin2{\ensuremath{(g-2)_\mu}}

\definecolor{Orange}{named}{orange}
\definecolor{Purple}{named}{purple}
\definecolor{Lightblue}{cmyk}{0.9,0.1,0.1,0.3}
\definecolor{dgelborange}{cmyk}{0.,0.3,0.5, 0.}
\definecolor{Lila}{rgb}{0.5,0.,1}
\definecolor{Darkgreen}{rgb}{0.,.7,0.2}